\documentclass[aps,prb,preprint,groupedaddress,floatfix]{revtex4}
\usepackage{graphicx,amssymb}
\begin{document}

\title{Free energy calculations along entropic pathways III. Nucleation of capillary bridges and bubbles.} 
\author{Caroline Desgranges and Jerome Delhommelle}
\affiliation{Department of Chemistry, University of North Dakota, Grand Forks ND 58202}
\date{\today}

\begin{abstract}
Using molecular simulation, we analyze the capillary condensation and evaporation processes for Argon confined in a cylindrical nanopore. For this purpose, we define the entropy of the adsorbed fluid as a reaction coordinate and determine the free energy associated with both processes along entropic pathways. For capillary condensation, we identify a complex free energy profile resulting from the multi-stage nature of this phenomenon. We find capillary condensation to proceed through the nucleation of a liquid bridge across the nanopore, followed by its expansion throughout the pore to give rise to the stable phase of high density. In the case of capillary evaporation, the free energy profile along the entropy pathway also exhibits different regimes, corresponding to the initial destabilization of the layered structure of the fluid followed by the formation, and subsequent expansion, of a bubble across the nanopore.
\end{abstract}

\maketitle

\section{Introduction}
The stability of liquid bridges between solid surfaces, as well as the mechanism by which they form, is a central phenomenon in interface science and adhesion~\cite{derjaguin1992theory,coasne2004grand,saugey2005nucleation,major2006viscous,casanova2007direct,puibasset2008monte,edison2013dynamics}. The formation of these bridges, as well as the nucleation of bubbles, actually occurs in a wide range of systems that span many length scales, from colloidal systems to nanosized capillaries~\cite{szoszkiewicz2005nucleation,bruschi2010capillary,greiner2010local}. This has been shown e.g. by atomic force microscopy and surface force apparatus experiments, which have revealed how liquid contacts can form and snap off, paving the way for applications in nanotribology and nanolithography~\cite{lin1994study,willett2000capillary,gogotsi2001situ,he2001critical,heuberger2001density,patel2002stability,maeda2002nanoscale,jang2003capillary,weeks2005direct,berim2008nanodrop}. In recent years, the use of molecular modeling and simulation to provide a molecular level understanding of this phenomenon has drawn considerable interest~\cite{thommes2006adsorption,fortini2006phase,horikawa2011capillary,monson2012understanding,gommes2012adsorption,mszr2013adsorption,zeng2014condensation,hiratsuka2016mechanism}. Density functional theory calculations~\cite{restagno2000metastability,talanquer2001nucleation,ustinov2005modeling,men2009nucleation}, as well as molecular dynamics~\cite{zhang1994phase,gelb1997liquid,landman1998nanotribological,leung2000dynamics,yasuoka2000molecular} and Monte Carlo simulations~\cite{liu1991wetting,gac1994influence,gelb1999phase,bock1999phase,bolhuis2000transition,stroud2001capillary,liu2006monte,mota2007simplified,winkler2010capillary,nguyen2011monte,gor2012capillary,siderius2013use,van2015mechanism} have started to shed light on the formation of solid-like and liquid-like junctions in nanopores and on the dynamics of cavitation between solid surfaces. In particular, Neimark and coworkers~\cite{kornev2002capillary,neimark2003bridging,vishnyakov2003monte,vishnyakov2003nucleation,neimark2005monte,neimark2005birth,neimark2005vapor} have examined the capillary condensation process in cylindrical nanopores and have shown that the scenario proposed by Everett and Haynes~\cite{everett1972model} for a macroscopic capillary could also extend to nanosized capillaries. This suggested a very complex and intriguing pathway for the capillary condensation process, involving a series of structural changes in the adsorbed fluid and the formation of a liquid bridge across the pore section as an intermediate towards the high density adsorbed phase.

Condensation and evaporation processes are challenging processes to model and simulate. This is due to the large free energy barrier of nucleation that the system has to overcome to complete the phase transition~\cite{oxtoby1988nonclassical,ten1998numerical,ten1999numerical,shen1999computational,desgranges2011role,patel2011quantifying,desgranges2014unraveling,lauricella2015clathrate}. Simulating the pathways underlying condensation and evaporation in nanoscale capillaries requires suitable simulation techniques that allow for the sampling of these rare events by driving the formation of the new phase. In this work, we focus on elucidating both the capillary condensation and evaporation pathways and on shedding light on the relation between the structure of the nanoconfined fluid and its entropy. For this purpose, we extend the recently developed $\mu VT-S$ simulation method~\cite{FS1,FS2} to study these two phenomena. The $\mu VT-S$ approach is implemented within the grand-canonical ensemble, in which the chemical potential $\mu$, the temperature $T$ and the volume $V$ of the system are held fixed. As shown in the case of the nucleation of liquid droplets for single component systems~\cite{FS1} and binary mixtures~\cite{FS2}, this method provides a direct way to calculate the entropy of the system throughout the nucleation process and, therefore, to use it as a reaction coordinate for such activated processes. In this work, we carry out  $\mu VT-S$ simulations with the aim of (i) simulating the adsorption and desorption of the fluid in the grand-canonical statistical ensemble, that is especially suited to study adsorption phenomena, and (ii) using the entropy of the confined fluid as the reaction coordinate to elucidate the two pathways underlying nanoscale capillary condensation and evaporation. To achieve this, we impose a target value for the entropy of the system, through an umbrella sampling bias potential~\cite{torrie1977nonphysical}. Then, by varying the value of the entropy of the system, we are able to sample the entire pathway connecting the two nanoconfined phases, starting from the metastable confined vapor and ending with the nanoconfined liquid in the case of capillary condensation. Doing so, we are able to identify the different intermediate structures, specifically bridges and bubbles, that form in the nanopores during the capillary condensation and evaporation processes.

The paper is organized as follows. We start by discussing how we extend the $\mu VT-S$ approach to study the condensation and evaporation processes of Argon in MCM-41 silica mesoporous molecular sieves. We also present the potential models used for the adsorbate and its interaction with the cylindrical nanopore, and provide the technical details for the simulations. We then examine the capillary condensation process for Argon in the nanopore and determine the free energy barrier associated with the process. We carry out a detailed structural analysis to identify the formation of the liquid bridge along this pathway. We also discuss the capillary evaporation process that occurs during the desorption of the confined fluid and apply the same analyses as for the reverse pathway. We finally draw our main conclusions in the last section.

\section{Models}
We study the capillary condensation and evaporation of Argon in a cylindrical pore of 10 atomic diameters. The geometry is defined as follows. The cylindrical pore is aligned with the $z$-axis, and we define the length, or lateral dimension, as $L_z$. Simulations are carried out at the boiling point for Argon ($T=87.3$~K) within a nanopore found in MCM-41 silica mesoporous molecular sieves~\cite{vishnyakov2003nucleation}. We model Argon with a Lennard-Jones potential and take the following parameters for the exclusion diameter $\sigma=3.4$~\AA~ and for the potential well-depth $\epsilon/k_B=119.8$~K. The interactions between the adsorbed Argon atoms and the nanopore are modeled with the following functional form~\cite{neimark2003bridging,vishnyakov2003monte,vishnyakov2003nucleation,neimark2005monte,neimark2005birth,neimark2005vapor}
\begin{equation}
\begin{array}{ll}
U_{sf}(r,R)=&\\
\pi^2 \rho_s \epsilon_{sf}\sigma_{sf}^2 \left \{ {63 \over 32} \left[ {{R-r} \over \sigma_{sf}} \left( 1 + {r \over R} \right) \right]^{-10} F \left[ {-9 \over 2}, {-9 \over 2};1; \left( {r \over R} \right)^2 \right] -3 \left[ {{R-r} \over \sigma_{sf}} \left( 1+ {r \over R} \right) \right]^{-4} F \left[ {-3 \over 2}, {-3 \over 2};1; \left( {r \over R} \right)^2 \right] \right \} &\\
\end{array}
\end{equation} 
in which $r$ is the radial coordinate of the Ar atom adsorbed in the pore, $R$ is the pore radius (here $5 \sigma$), $\rho_s$ is the surface density of adsorption centers and $F(\alpha,\beta;\gamma;\delta)$ is the hypergeometric series. The parameters for the solid-fluid interactions were taken as $\rho_s \epsilon_{sf}=2253$~K/nm$^2$ and $\sigma_{sf}=3.17$~\AA. This functional form accurately models the interaction between adsorbate and the structureless cylindrical layer of adsorption centers on the pore wall as shown by Ravikovitch {\it et al.}~\cite{ravikovitch1997evaluation}. Following Vishnyakov and Neimark~\cite{vishnyakov2003nucleation}, we carry out simulations of capillary condensation and evaporation in nanopores with a long lateral dimension $L_z=30 \sigma$ to allow for the sampling of symmetry breaking configurations containing bubbles and liquid bridges. Periodic boundary conditions are applied along this lateral direction $z$. We also calculate explicitly the interactions between Argon atoms up to a distance of $5 \sigma$ and neglect the fluid-fluid interactions beyond that cutoff distance. Finally, in the rest of this work, we use the conventional set of reduced units~\cite{Allen}, with respect to the Lennard-Jones parameters of the fluid.

\section{Simulation method}
To span the entire pathway underlying the capillary condensation and evaporation processes, we carry out a series of $\mu VT-S$ simulations and gradually vary the entropy of the system to drive the phase transition within the confined fluid. We briefly outline here the principles of the $\mu VT-S$ method (more details may be found in previous work~\cite{FS1,FS2}). $\mu VT-S$ simulations are carried out in the grand-canonical ensemble $(\mu,V,T)$, i.e. at constant chemical potential $\mu$, temperature $T$ and volume $V$. This statistical ensemble is especially well suited to study adsorption phenomena, since the number of atoms $N$ adsorbed in the nanopore is allowed to vary in the $(\mu,V,T)$ ensemble. Thus, this ensemble mimics what is observed experimentally during the adsorption/desorption processes. This ensemble has also the advantage of providing a direct way of estimating the entropy of the confined fluid. Taking the total Helmoltz free energy $A=U-TS$ of the confined fluid to be equal to $N\mu$, we obtain the following equation for the entropy of the adsorbed fluid $S_{loading}$
\begin{equation}
S_{loading}= {{U - \mu N} \over T}
\label{loading}
\end{equation}
In this equation, $S_{loading}$ denotes the total entropy of the system and, as such, increases with the total number of atoms $N$ adsorbed in the nanopore.  We add that the $pV$ term is omitted in Eq.~\ref{loading} since it is negligible when compared to the other terms ($pV$ actually accounts for less than $0.5$~\% of the lowest values sampled for $TS_{loading}$). This approximation is similar to the one made by Waghe, Rasaiah and Hummer in their calculation of the entropy of water adsorbed in carbon nanotubes~\cite{Waghe}. To sample the transition pathway, we use $S_{loading}$ as the reaction coordinate and drive the nucleation event through an umbrella sampling approach~\cite{torrie1977nonphysical}. For this purpose, we define the following bias energy:
\begin{equation}
U_{b}= {1\over 2} k (S_{loading}-S_0)^2 
\label{biasfunc}   
\end{equation}
in which $S_0$ is the target value for the total entropy. This bias energy is added to the total potential energy of the system, which appears in the Metropolis criteria used in the Monte Carlo simulations. In practice, for capillary condensation, we perform a series of $\mu VT-S$ simulations with increasing values for the target entropy $S_0$. This promotes the uptake of additional Ar atoms since the total entropy of the system is a function of $N$. Similarly, the pathway for the desorption process is sampled by carrying out successive $\mu VT-S$ simulations, with decreasing values for the target entropies $S_0$. 

Previous work~\cite{peterson1987phase} has shown that the conditions for liquid-vapor equilibrium in the pore are obtained for $\mu =-10.53 \epsilon$. To observe the capillary condensation/evaporation processes, the chemical potential must be close enough to the chemical potential at equilibrium. More specifically, to simulate the capillary condensation process, we carry out $\mu VT-S$ simulations for a chemical potential of $\mu=-10.48\epsilon$, while for the evaporation process, we use a chemical potential of $\mu=-10.54\epsilon$. These two values for $\mu$ show that our findings are consistent with the results from Peterson and Gubbins~\cite{peterson1987phase}. As expected, these two values of $\mu$ bracket the estimate of Peterson and Gubbins for the chemical potential at the liquid-vapor equilibrium, with $\mu$ for the capillary condensation process being slightly above the chemical potential at coexistence and $\mu$ for the capillary evaporation process being slightly below. A total of $35$ umbrella sampling windows are performed to connect the vapor and liquid phases for the confined fluid and to sample the entire condensation and evaporation pathways. For each window, we first carry out an equilibration run of $1 \times 10^8$ MC steps to allow the system to relax towards the target value for the entropy. We then perform a production run of $2 \times 10^8$ MC steps, during which averages are collected collect and structural analyses are run.

\section{Results and Discussion}

\begin{figure}
\begin{center}
\includegraphics*[width=8cm]{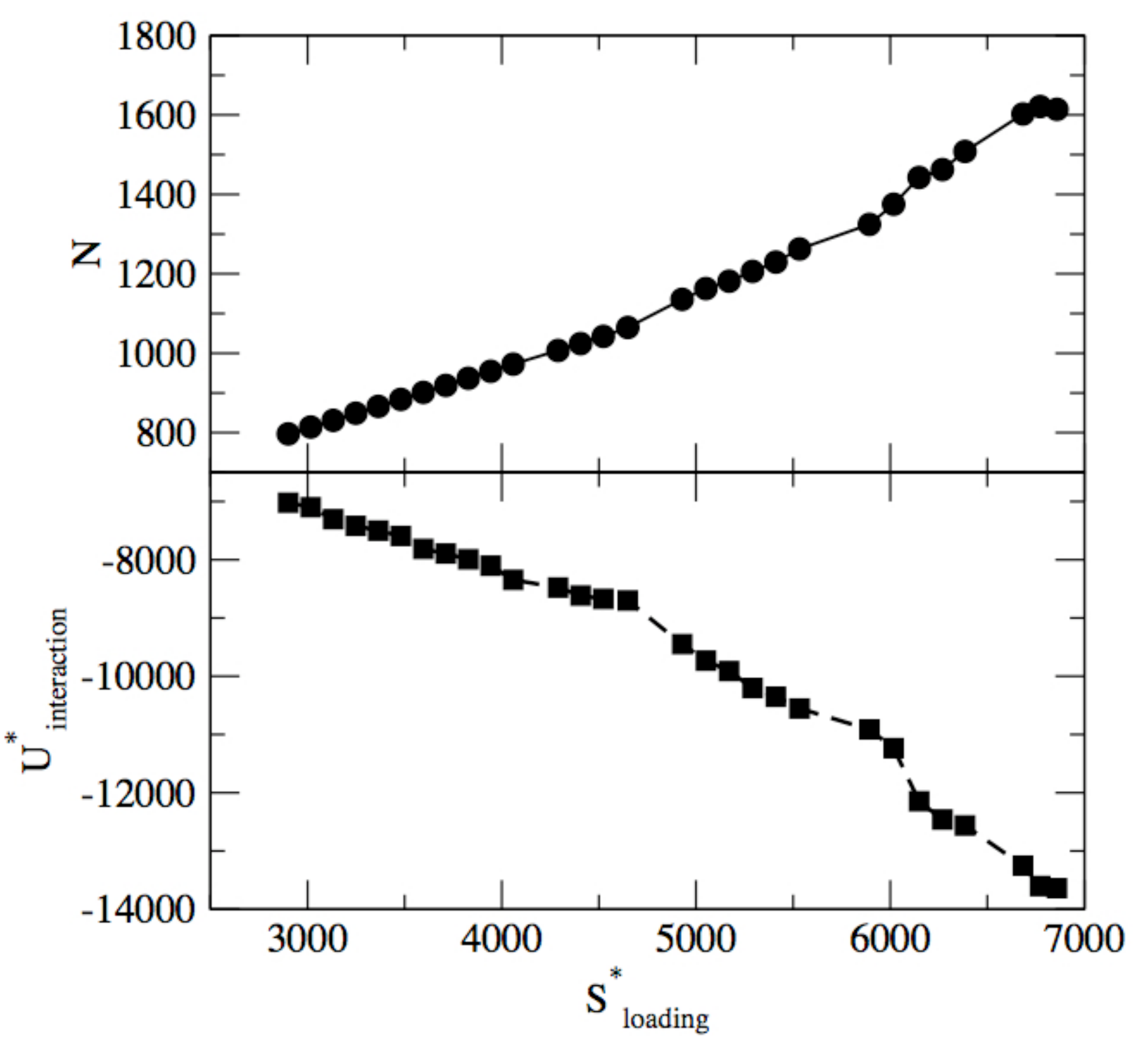}
\end{center}
\caption{Capillary condensation ($\mu=-10.48\epsilon$): variation of the number of atoms adsorbed $N$ (top) and of the interaction energy for the confined fluid $U^*_{potential}$ (bottom) as a function of the total entropy of the adsorbed fluid $S^*_{loading}$.}
\label{Fig1}
\end{figure}

We start by discussing the results obtained for the capillary condensation process. To sample the entropic pathway for this process, we carry out a series of $\mu VT-S$ simulations at $\mu=-10.48\epsilon$. We gradually increase the value for the target entropy, as we go from one umbrella sampling window to another, and, as a result, increase $S_{loading}$ for the confined fluid. We show in Fig.~\ref{Fig1} how the system responds to this increase in the entropy of the adsorbed fluid. The top panel of Fig.~\ref{Fig1} consists of a plot of the number of Argon atoms $N$ adsorbed in the nanopore as a function of the reduced entropy of the adsorbed fluid (noted as $S^*_{loading}$). We observe the following trends. First, we find that $N$ increases with $S^*_{loading}$, starting from about $900$~Ar atoms adsorbed for $S^*_{loading}=3300$, and reaches a value of about $1600$~Ar atoms for $S^*_{loading}=6800$. This plot shows that the successive $\mu VT-S$ simulations, with increasing values for the total entropy of the system, allow to simulate the phase transition process from a metastable phase of low density to the stable phase of high density. The bottom panel of Fig.~\ref{Fig1} shows the corresponding variation of the interaction energy within the confined fluid during the $\mu VT-S$ simulations. Overall, we observe a decrease in the interaction energy as $S^*_{loading}$ increases. This behavior is consistent with the increase in the amount of $Ar$ adsorbed in the cylindrical nanopore observed when $S^*_{loading}$ increases.

\begin{figure}
\begin{center}
\includegraphics*[width=8cm]{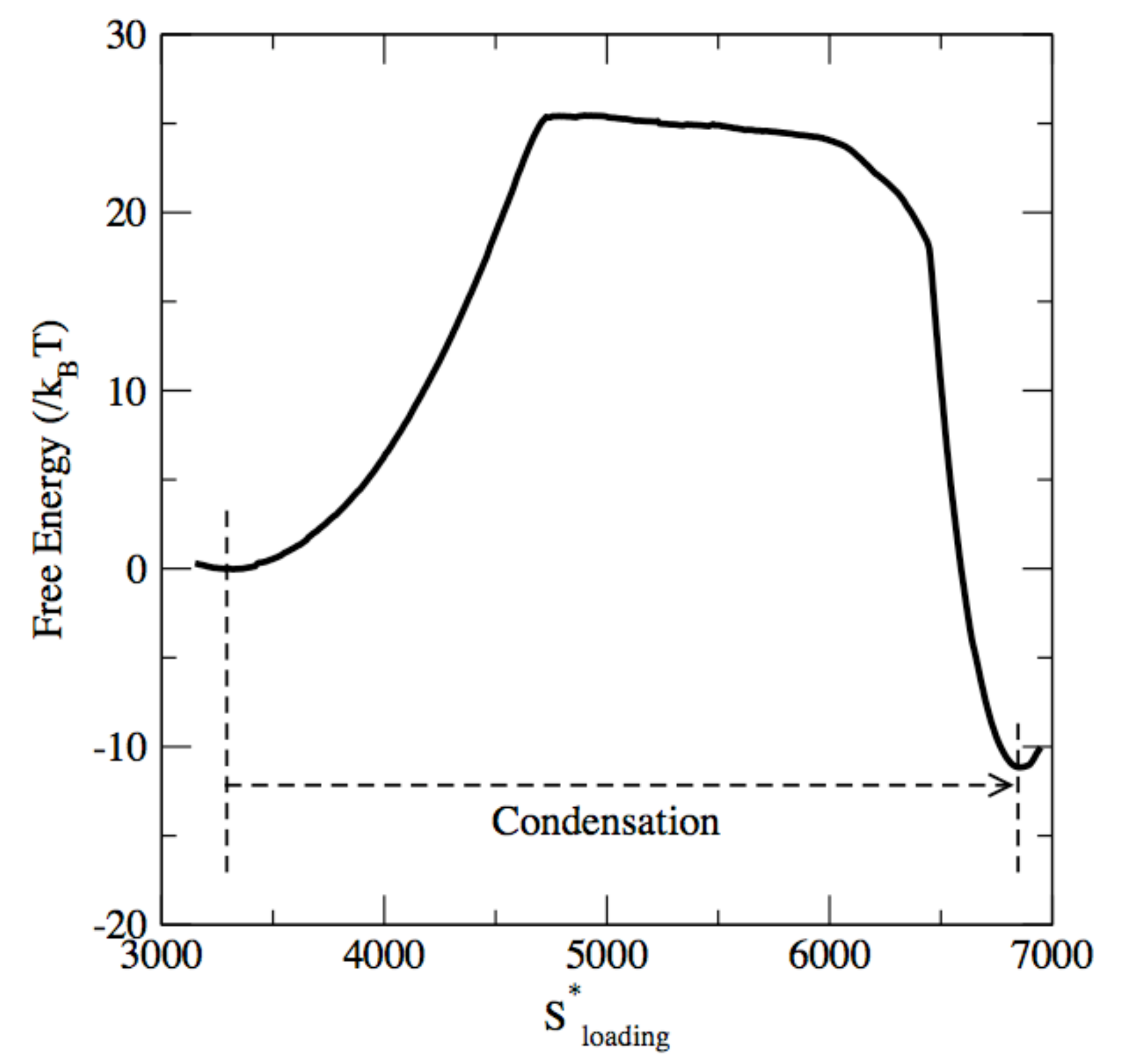}
\end{center}
\caption{Free energy profile for the capillary condensation at $\mu=-10.48\epsilon$. The origin for the energy is set to $0$ for the starting point (metastable phase of low density - see the local minimum on the left of the plot). After condensation, the system reaches a free energy minimum corresponding to the stable phase of high density (see the minimum on the right of the plot).}
\label{Fig2}
\end{figure}

We plot in Fig.~\ref{Fig2} the free energy profile obtained for the capillary condensation process. The starting point for the $\mu VT-S$ simulations is a metastable phase of low density adsorbed in the nanopore, with a reduced entropy of $S^*_{loading}=3300$. As shown on the left of Fig.~\ref{Fig2}, the free energy profile exhibits a local minimum for this metastable phase. For convenience, we have chosen to assign the origin for the free energy to this local minimum. Several regimes can be identified in this plot of the free energy as a function of the entropy of the adsorbed phase. First, the free energy profile exhibits a steep increase, over entropies ranging from $S^*_{loading}=3300$ to $S^*_{loading}=4800$. This increase in free energy corresponds to the steady increase in $N$, and accordingly to the steady decrease of $U^*_{interaction}$, over the same range of entropies. Then, for $S^*_{loading}=4800$ to $S^*_{loading}=6000$, the free energy profile exhibits a second regime which can be characterized as an almost flat top, that very slowly decreases with $S^*_{loading}$. This second regime is associated with changes in both the slopes for the variations of $N$ and $U^*_{interaction}$ in Fig.~\ref{Fig1}. Then, for $S^*_{loading}=6000$ to $S^*_{loading}=6850$, we observe a third regime for the free energy profile and a steady decrease in free energy. During this stage, the system completes the phase transition towards the stable phase of high density. As expected, we obtain a free energy for the stable phase that is below, by $11$~$k_BT$, the free energy of the metastable phase of low density. This establishes that the end point for the $\mu VT-S$ simulations is indeed the stable phase. As for the previous two regimes in the free energy profile, we find that the decrease for the free energy, observed during the third regime, is connected to sharp changes in the variations of both $N$ and $U^*_{interaction}$ for the adsorbed phase. This indicates that significant structural changes take place within the fluid confined in the nanopore.

\begin{figure}
\begin{center}
\includegraphics*[width=8cm]{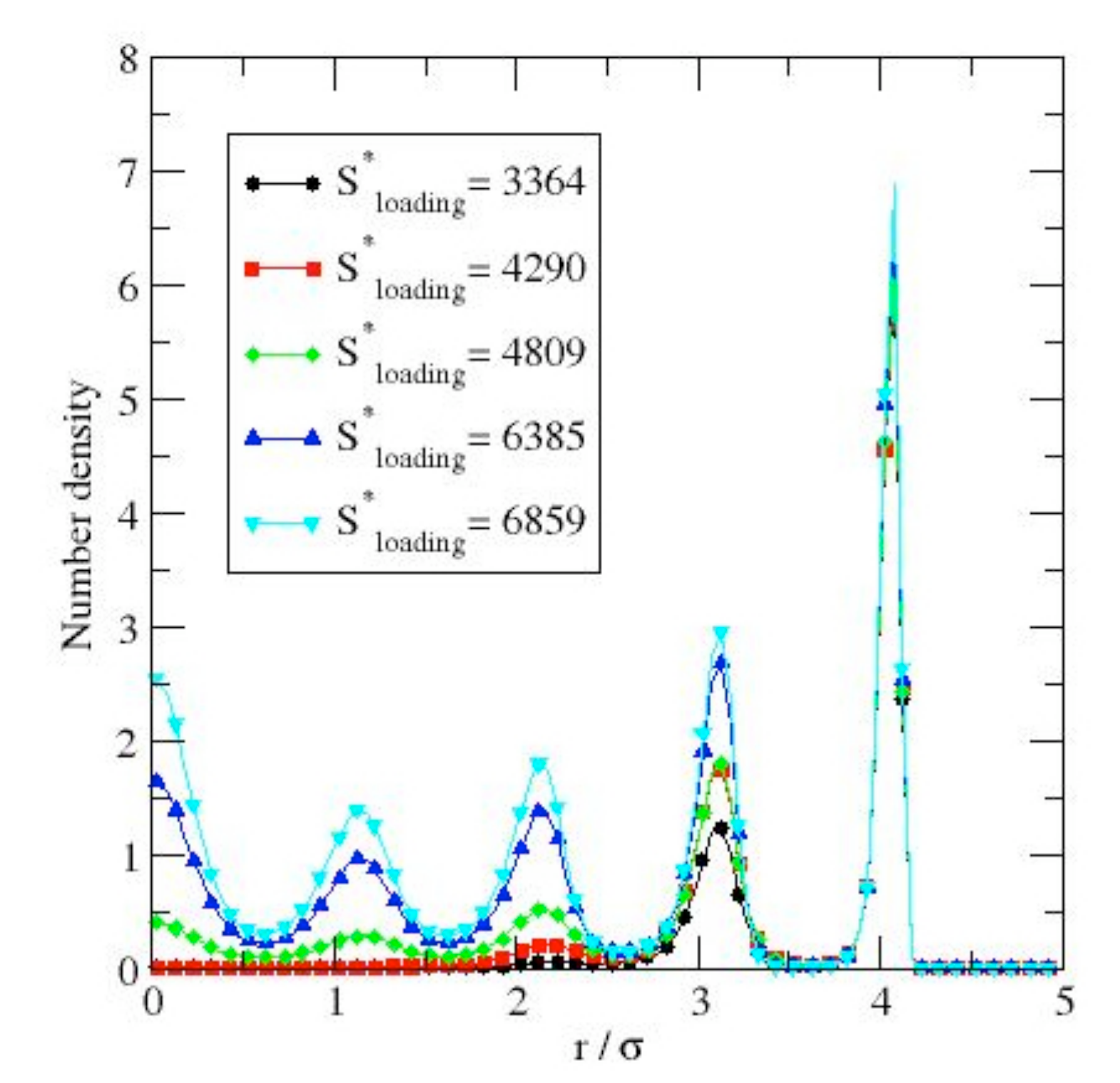}
\end{center}
\caption{Capillary condensation ($\mu=-10.48$): density profiles across the nanopore ($r=0$ denotes the center of the pore) for increasing values of $S^*_{loading}$. The profile in black ($S^*_{loading}=3364$) is for the metastable phase of low density, while the profile in cyan ($S^*_{loading}=6859$) corresponds to the stable phase of high density.}
\label{Fig3}
\end{figure}

To elucidate the mechanism underlying capillary condensation, we carry out a series of structural analyses. We begin by determining the density profiles for the adsorbed fluid across the nanopore. The results obtained during several of the $\mu VT-S$ simulations are shown in Fig.~\ref{Fig3}. At the start of the capillary condensation process, i.e. for the metastable phase of low density, the adsorbed fluid mostly consists of two layers close to the wall, as shown by the two peaks on the density profile for radii of about $4$~$\sigma$ (fluid layer closest to the wall) and $3$~$\sigma$ (second layer). As $S^*_{loading}$ increases, $Ar$ atoms start to populate the center of the nanopore, and a third peak for a radius around $2$~$\sigma$ develops (see the density profile in red for $S^*_{loading}=4290$). We detect the formation of a fourth and then of a fifth peak when the system reaches the top of the free energy barrier (see density profile in green). Then, as $S^*_{loading}$ further increases, the height of the peaks associated with the innermost fluid layers continue to increase, until the nanopore has completely filled and the system reaches the stable phase of high density. 

\begin{figure}
\begin{center}
\includegraphics*[width=7.8cm]{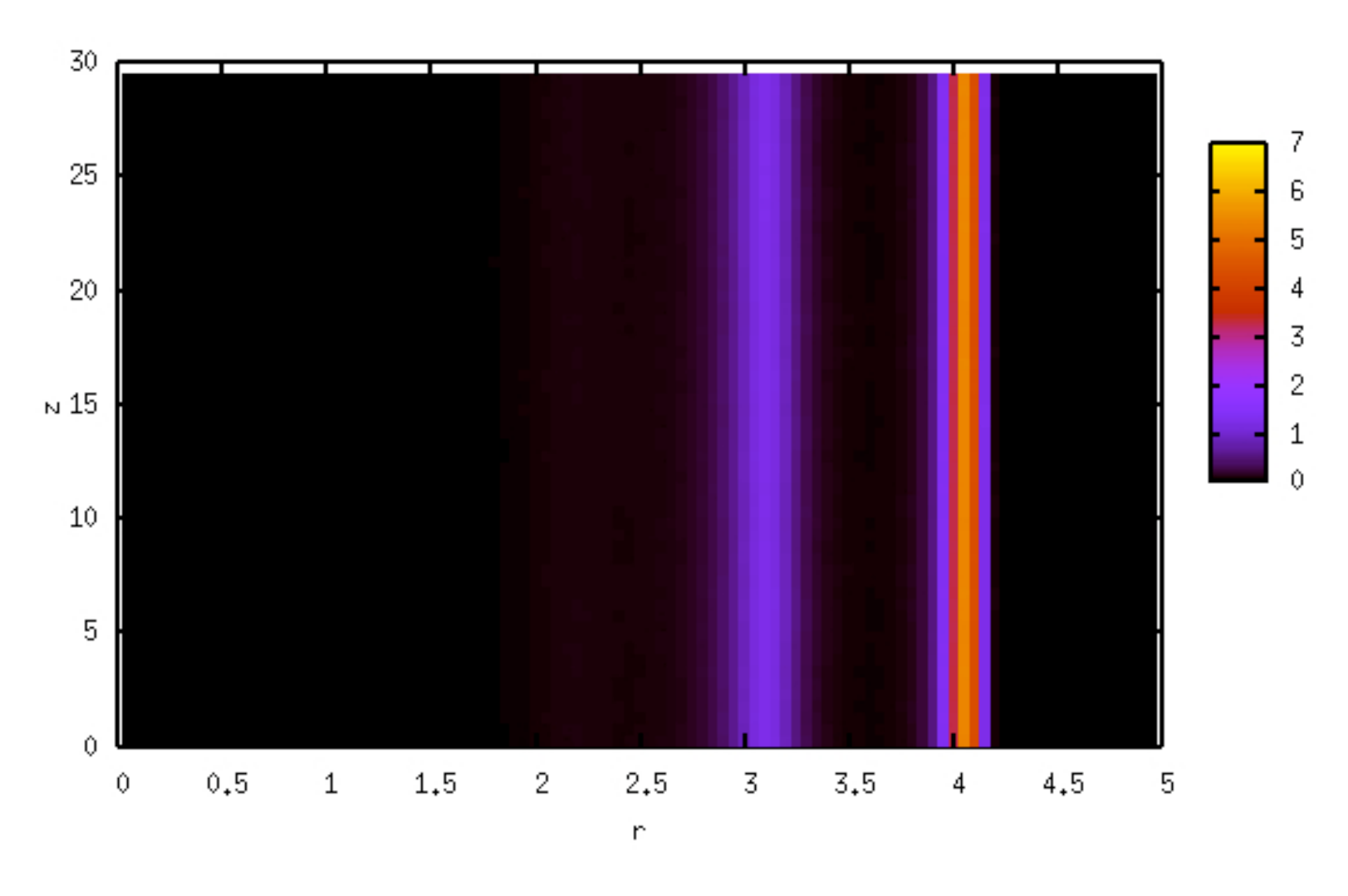}
\includegraphics*[width=7.8cm]{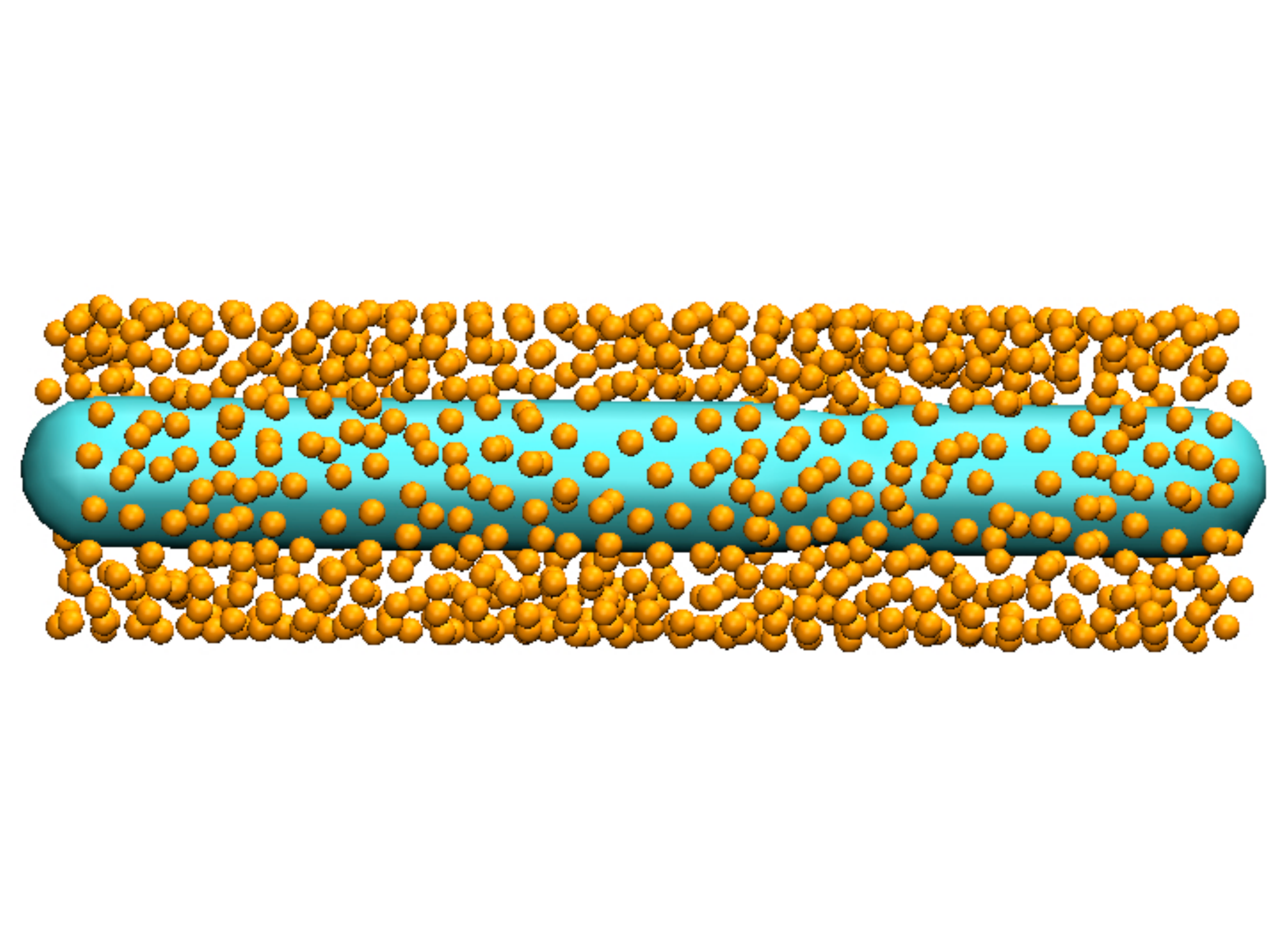}(a)
\includegraphics*[width=7.8cm]{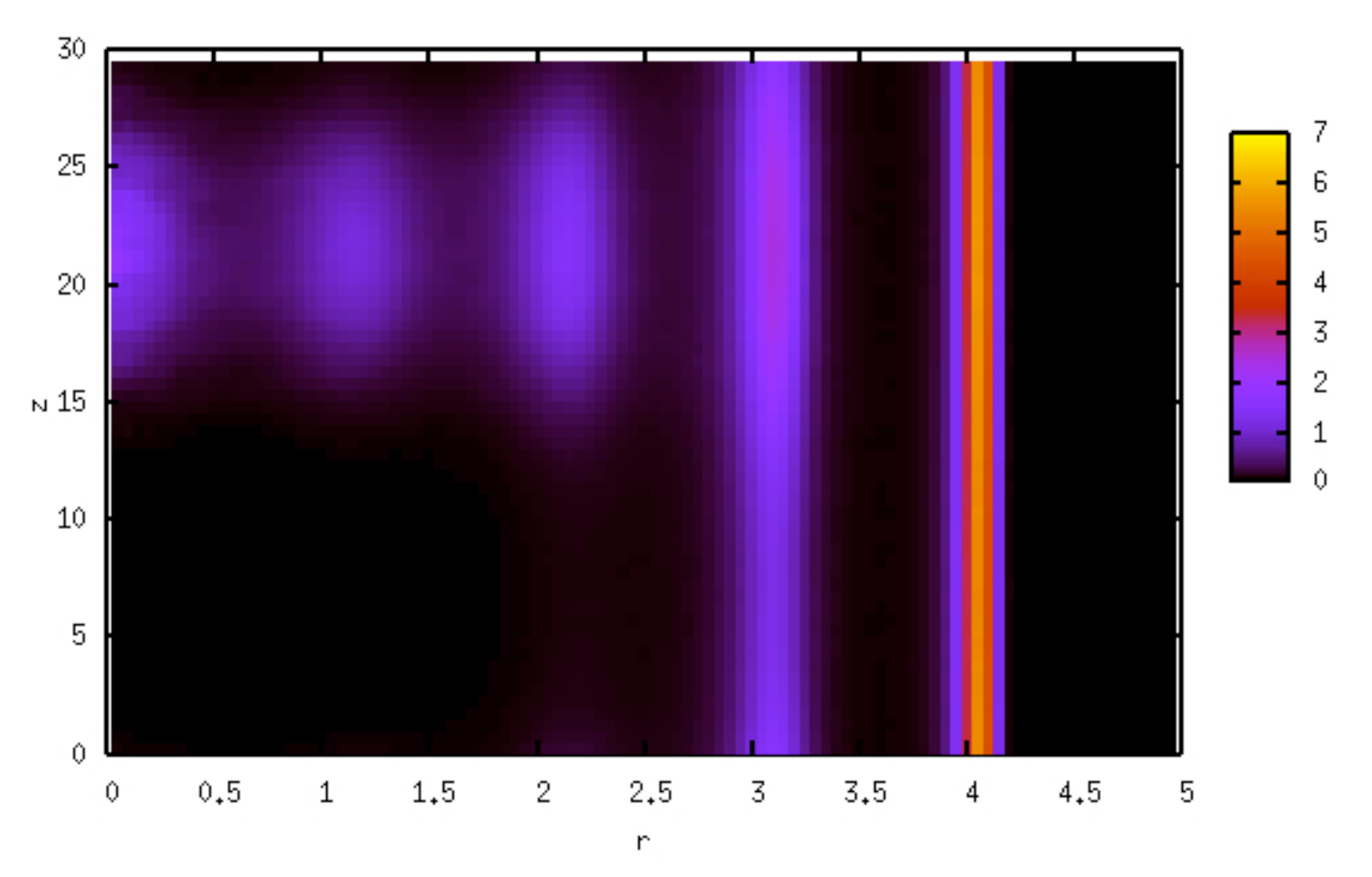}
\includegraphics*[width=7.8cm]{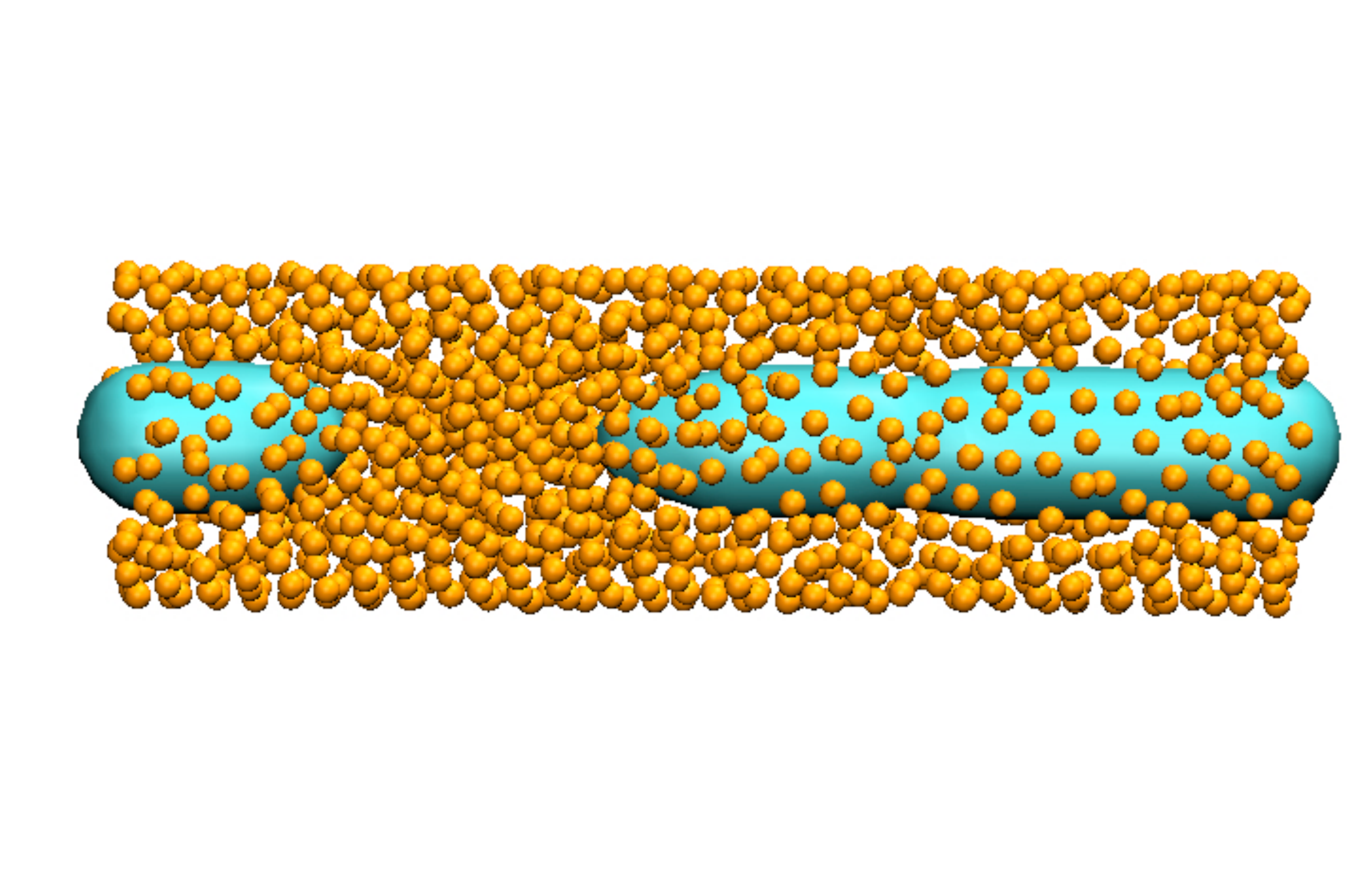}(b)
\includegraphics*[width=7.8cm]{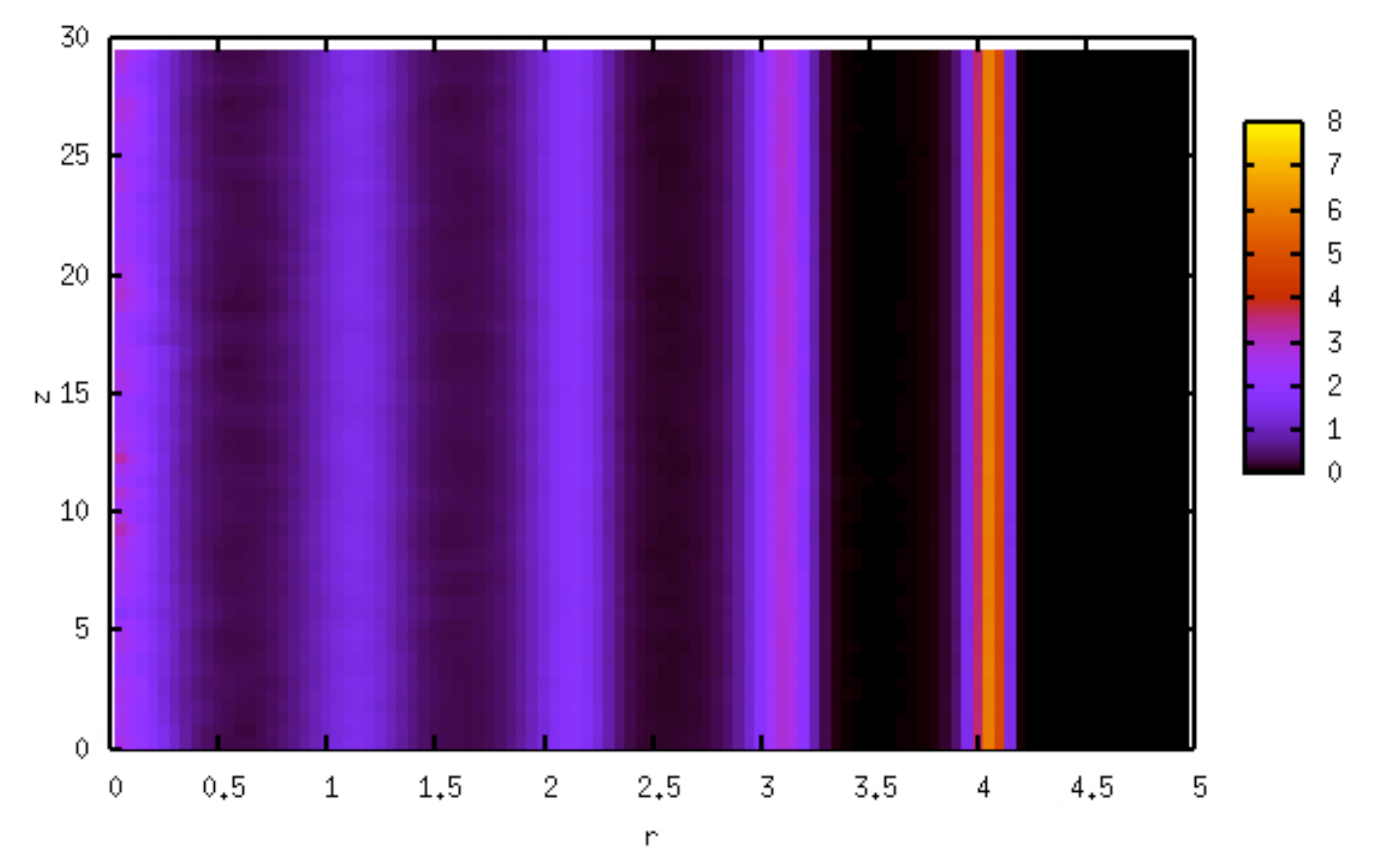}
\includegraphics*[width=7.8cm]{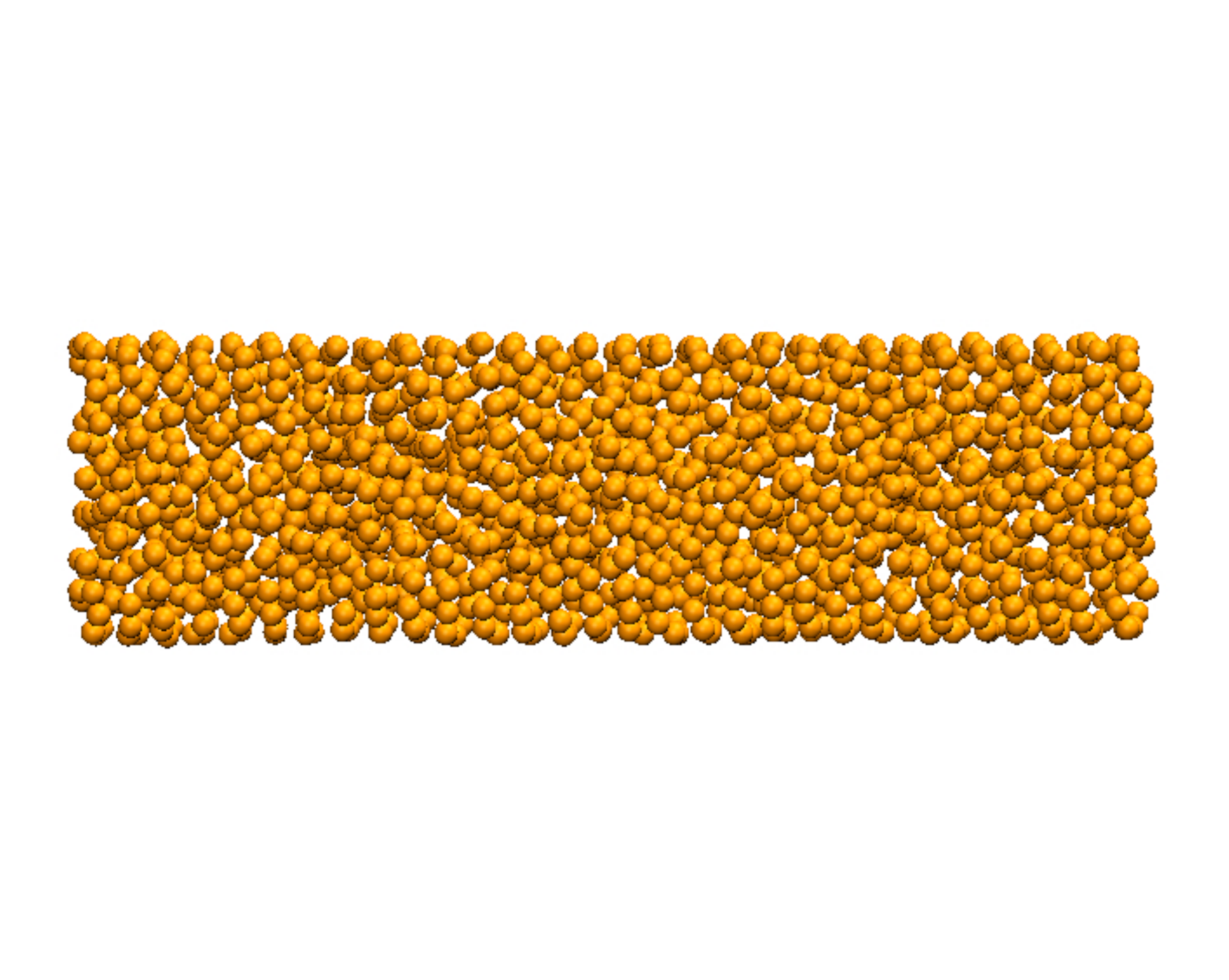}(c)
\end{center}
\caption{Mechanism underlying capillary condensation. Radial density profile along the side ($z$) of the nanopore for increasing values of the entropy $S^*_{loading}=3364$ in (a), $S^*_{loading}=4929$ in (b) and $S^*_{loading}=6859$ in (c), together with the corresponding snapshots. In the snapshots, $Ar$ atoms are shown as orange spheres, while the regions of low density are highlighted in cyan. The bright spots on the left of the density profile around $z=25$ indicate the formation of the liquid bridge in (b), as shown by the discontinuity of the cyan region in the corresponding snapshot.}
\label{Fig4}
\end{figure}

We now more closely look at the radial density profiles of Fig.~\ref{Fig3} to determine why, for intermediate values of $S^*_{loading}$, we find lower peaks found for the innermost fluid layers. This prompts the following question. Are these lower peaks associated with a uniform distribution of Ar atoms along the side of the tube (along the $z$ direction)? Alternatively, do these lower peaks signal the formation of a liquid contact within the nanotube? To answer this question, we need to determine how the distribution of atoms evolves along the side of the cylindrical nanopore, i.e. along the $z$ axis. For this purpose, we refine our structural analysis by computing the density profiles against both the radius and the $z$ coordinate and show how the radial density varies along the nanopore. We show in Fig.~\ref{Fig4}(a) the results obtained for the metastable phase of low density. The plot for the density profile shows that the atoms adsorbed are very predominantly located close to the wall. More specifically, the fluid is organized into two layers close to the wall as shown by the two bright lines (i.e. regions of high density) on the density profile of Fig.~\ref{Fig4}(a). This is consistent with the profiles of Fig.~\ref{Fig3} that showed the formation of two layers close to the wall. The additional information brought by Fig.~\ref{Fig4}(a) is the fact that $Ar$ atoms in these two layers are distributed uniformly along $z$. We also show in Fig.~\ref{Fig4}(a) a snapshot of a configuration of the system for the metastable phase of low density and highlight in cyan the region where the void dominates (we define void any region of space for which the reduced density is below a threshold value). We carry out the same analysis along the rest of the capillary condensation pathway. At the top of the free energy barrier, we obtain the density profile and snapshot shown in Fig.~\ref{Fig4}(b). Both show that, at this stage, the $Ar$ atoms located in the inner part of the nanopore are not uniformly distributed along z and that a liquid-like region has developed for a value of $z$ around $20$~$\sigma$ on the density profile. This can best be seen on the density profile with the onset, for $z=20$, of 3 bright spots for $r$ around $0$, $1$ and $2$, that indicate the formation of 3 partial liquid layers. The snapshot of a configuration of the system at the top of the free energy barrier (see Fig.~\ref{Fig4}(b)) confirms that a liquid bridge has developed across the nanopore for this value of $z=20$. As the capillary condensation further proceeds, the liquid bridge becomes wider and wider along the nanopore, resulting in a recess of the void region. Finally, as the system reaches the stable high density phase, we see that the structure of the adsorbed fluid exhibits a uniform structure with multiple layers (see Fig.~\ref{Fig4}(c)). The mechanism observed for capillary condensation is thus consistent with the scenario proposed by Everett and Haynes~\cite{everett1972model} for a macroscopic capillary and the results from gauge cell Monte Carlo simulations from Neimark {\it et al.}~\cite{vishnyakov2003nucleation}, which both pointed to a liquid bridge-mediated capillary condensation process. We add that the free energy barriers are of the same order as those previously obtained for the capillary condensation process by Vishnyakov and Neimark~\cite{vishnyakov2003nucleation} using the gauge cell method.

\begin{figure}
\begin{center}
\includegraphics*[width=8cm]{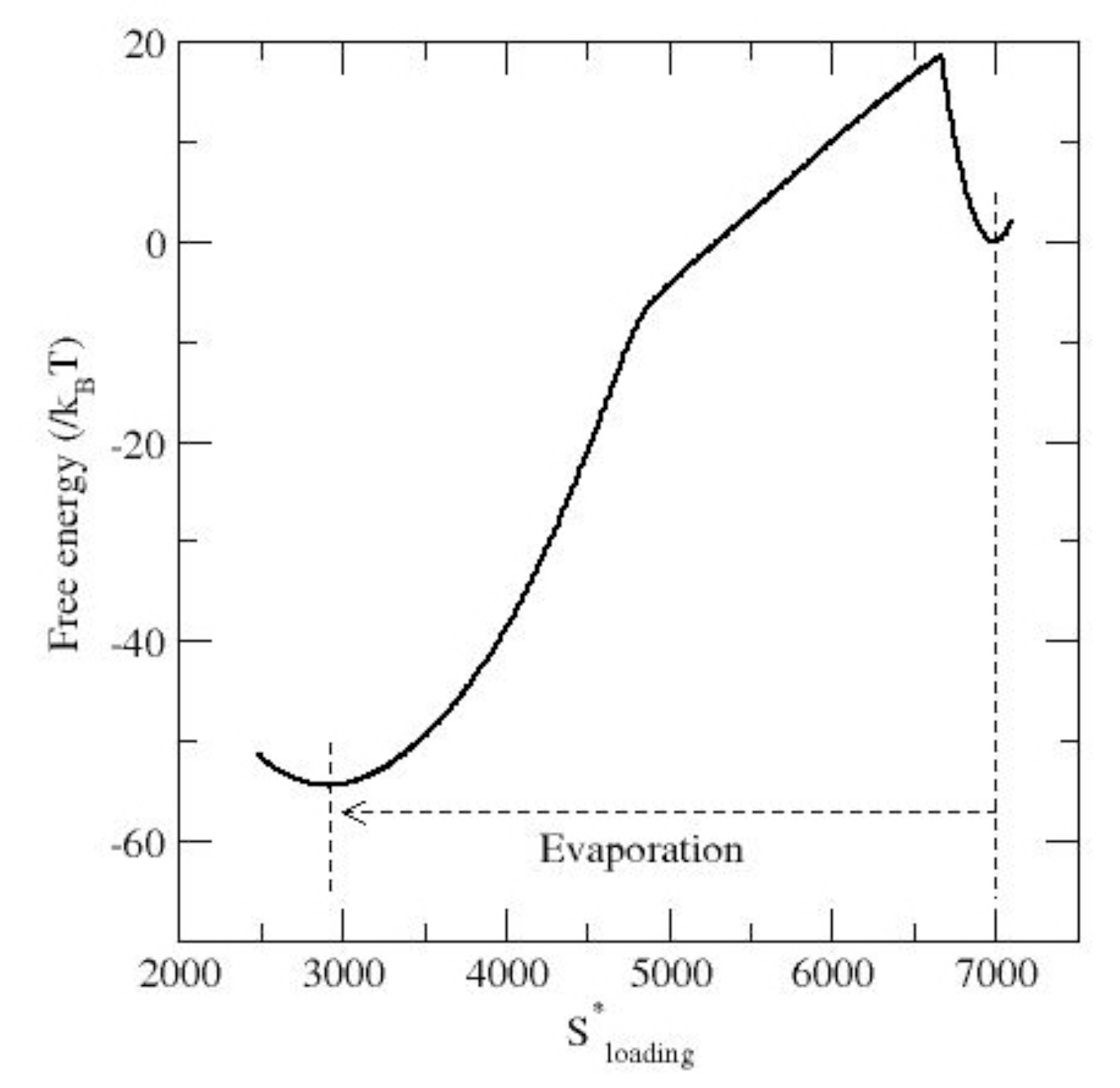}
\end{center}
\caption{Free energy profile for the capillary evaporation at $\mu=-10.54 \epsilon$. The origin for the energy is set to $0$ for the starting point (metastable phase of high density). After evaporation, the system reaches a free energy minimum corresponding to the stable phase of low density.}
\label{Fig5}
\end{figure}

We now turn to the results obtained for the capillary evaporation process ($\mu=-10.54 \epsilon$). We show in Fig.~\ref{Fig5} the free energy profile obtained as the system moves from a metastable phase of high density to the stable phase of low density. In line with our results for the capillary condensation process, we set the origin for the free energy axis to the free energy of the metastable phase. The main difference with the case of the condensation process, studied in the first part of the paper, is that we now carry out a series of $\mu VT-S$ simulations with decreasing values for the total entropy of the system. Therefore, $S^*_{loading}$ now varies in the opposite sense to the one followed for the condensation process. Looking at the free energy profile as a function of $S^*_{loading}$, we find three successive regimes for the variations of the free energy of the system. First, for $S^*_{loading}$ varying from $7000$ (metastable phase of high density) to $6500$ (top of the free energy barrier), we observe a steep increase in free energy. Then, for $S^*_{loading}$ between $6500$ and $4800$, the free energy profile exhibits a much more gradual behavior, as the free energy slowly decreases with the entropy of the system, in an almost linear fashion. The third regime is observed for lower values of $S^*_{loading}$ (between $4800$ and $2900$). For this range of entropies, the free energy decreases more rapidly with entropy, until the system reaches the stable phase of low density. At this point, the free energy of the system is $55$~$k_BT$ lower than the metastable state from which the evaporation pathway started. This confirms that the system has reached a thermodynamically stable phase at the end of the $\mu VT-S$ simulations.

\begin{figure}
\begin{center}
\includegraphics*[width=8cm]{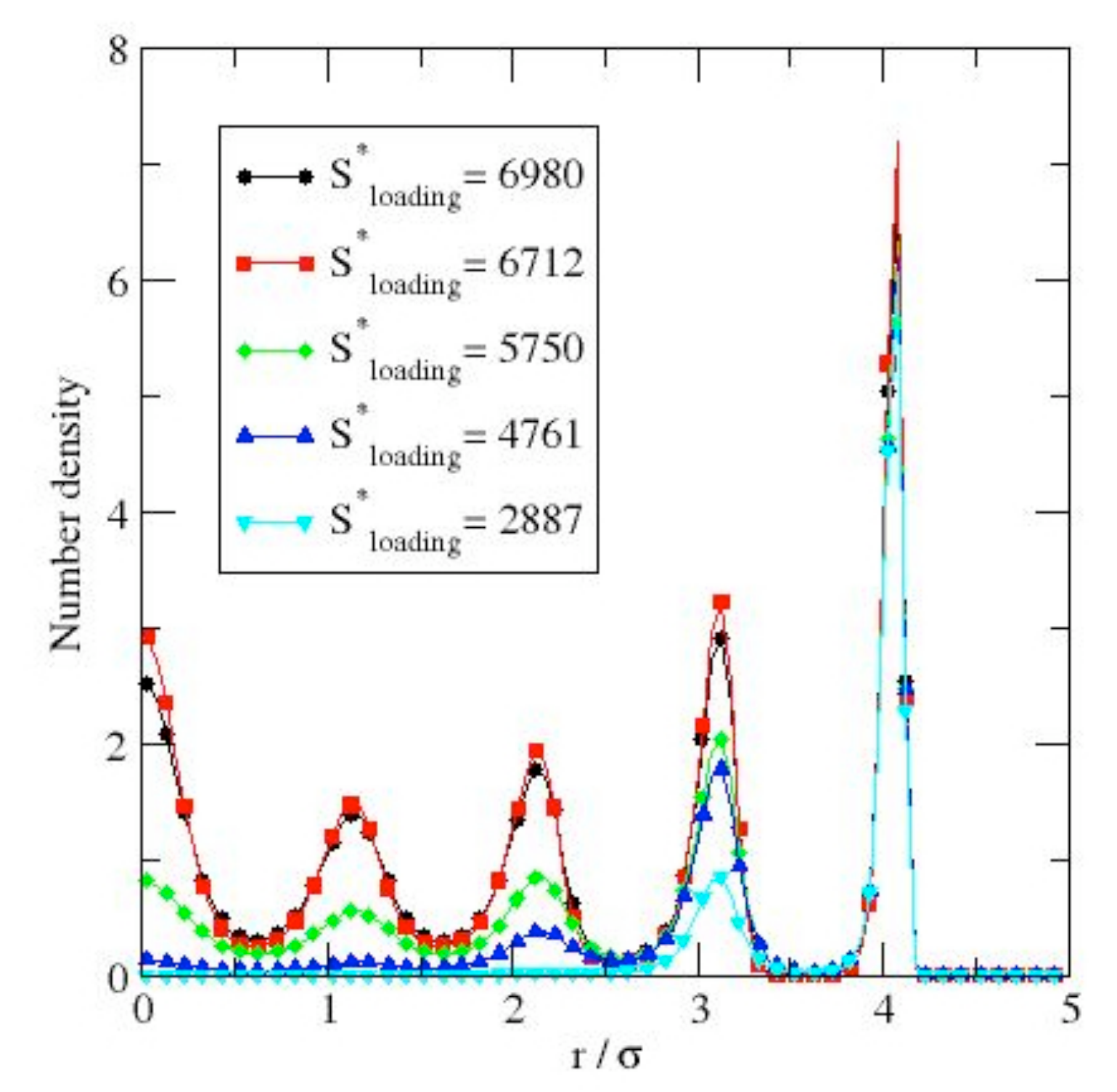}
\end{center}
\caption{Capillary evaporation ($\mu=-10.54 \epsilon$): density profiles across the nanopore ($r=0$ denotes the center of the pore) for decreasing values of $S^*_{loading}$. The profile in black ($S^*_{loading}=6980$) is for the metastable phase of high density, while the profile in cyan ($S^*_{loading}=2887$) corresponds to the stable phase of low density.}
\label{Fig6}
\end{figure}

We now analyze the structure of the adsorbed fluid along the evaporation pathway. Fig.~\ref{Fig6} shows the density profiles across the cylindrical nanopore for different values of the entropy. The starting point for the $\mu VT-S$ simulations displays the expected multi-peaked density profile. This density profile is consistent with the organization of the metastable phase of high density in multiple fluid layers within the nanopore ($S^*_{loading}=6980$). Close to the top of the free energy barrier ($S^*_{loading}=4761$), we observe a small, but noticeable, increase in the number of $Ar$ atoms adsorbed in the nanopore. This indicates that the capillary evaporation process starts with the adsorption of a few extra $Ar$ atoms, amounting to about $1-2\%$ of the overall number of $Ar$ atoms within the system. These extra $Ar$ atoms destabilize the organization within the nanopore and trigger the desorption process, as shown by the decrease in the amplitude of the peaks in the density profile that follows (see e.g. the density profile obtained for $S^*_{loading}=5750$). This leads to a decrease in the free energy as $S^*_{loading}$ further decreases. Then, at the end of the second regime found in the free energy plot, we observe that the density profile exhibits two main peaks, the third inner most layer showing a reduced fluid density ($S^*_{loading}=4761$). Finally, we recover the structure of the metastable phase of low density (see the density profile for at $S^*_{loading}=2887$), for which the $Ar$ atoms are very predominantly adsorbed close to the wall of the cylindrical nanopore.

\begin{figure}
\begin{center}
\includegraphics*[width=7cm]{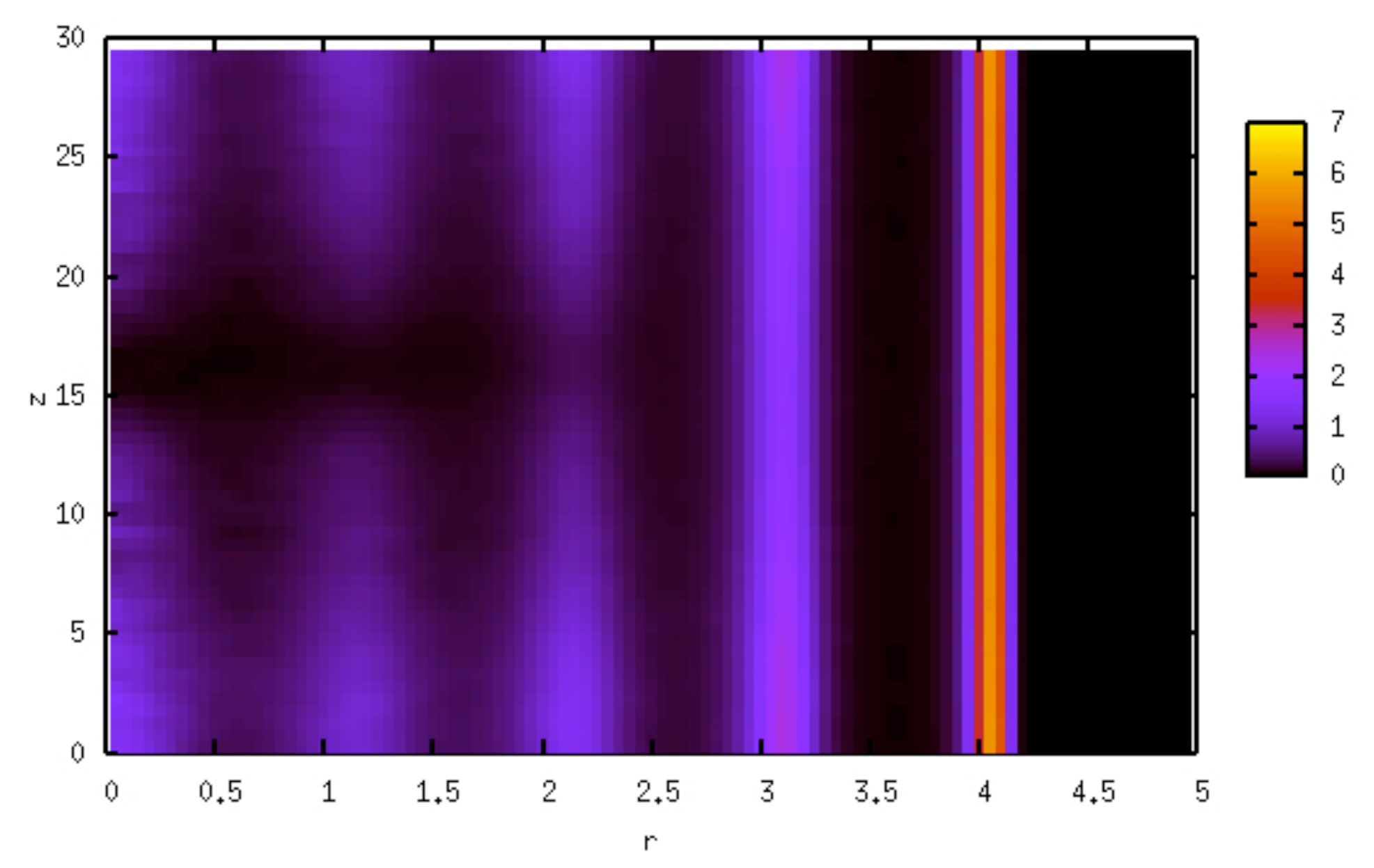}
\includegraphics*[width=7cm]{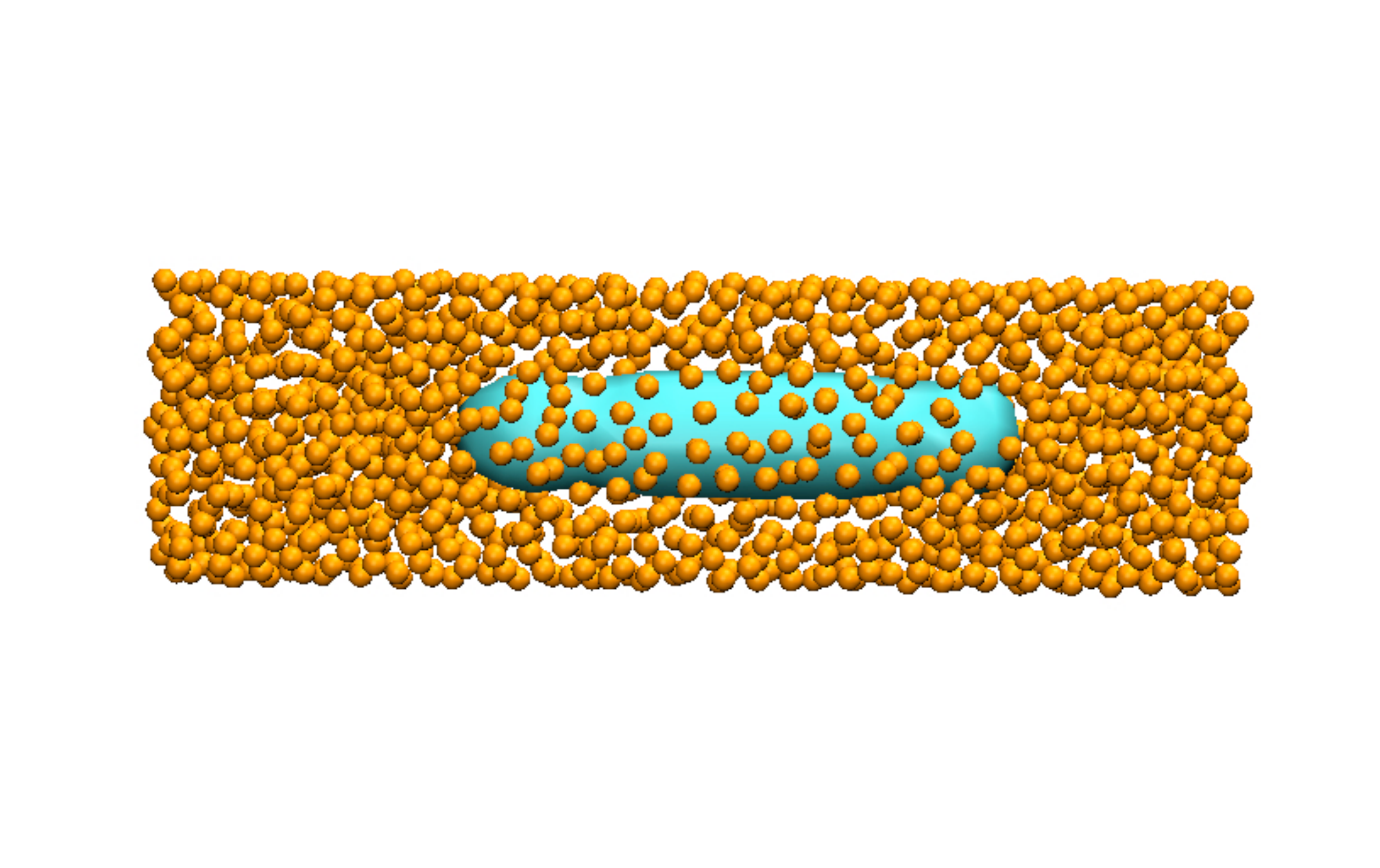}
\end{center}
\caption{Bubble formation during capillary evaporation. Radial density profile along the side ($z$) of the nanopore for $S^*_{loading}=5750$ and its corresponding snapshot. The darker spots on the left of the density profile around $z=15$ indicate the formation of the bubble, which is highlighted in cyan in the snapshot.}
\label{Fig7}
\end{figure}

We finally comment on the local structure within the fluid along the side of the nanopore. We observe that once the organization of the high density phase has been destabilized, capillary evaporation proceeds through the formation of a bubble within the nanopore. This can best be seen in Fig.~\ref{Fig7} for $S^*_{loading}=5750$. The density profile of Fig.~\ref{Fig7} shows that a dark spot develops for values of $z$ around 15 and that the innermost layers of the adsorbed fluid start to be depleted. Furthermore, this plot establishes that a non-uniform distribution of $Ar$ atoms has formed along the side of the nanopore. In this case, the dark spots on the density profile on the left of Fig.~\ref{Fig7}, as well as the reduction in the height of the peaks for the innermost layers in Fig.~\ref{Fig6}, are due to the formation of a bubble, as shown through the low density region indicated on the snapshot on the right of Fig.~\ref{Fig7}.

\section{Conclusions}
In this work, we carry out $\mu VT-S$ simulations to elucidate the capillary condensation and evaporation processes. Using the total entropy for the adsorbate, we are able to drive these processes along the entropic pathways underlying these processes. This allows us to shed light on the structural changes that occur within the confined fluid and give rise to the phase transition. Considering the example of Argon adsorbed in a smooth cylindrical nanopore typical of the MCM-41 silica adsorbent, we start by focusing on capillary condensation. Our simulations allow us to identify a complex free energy profile, corresponding to the successive stages in the condensation process. Our results show that capillary condensation from a metastable phase of low density starts with the nucleation of a liquid bridge within the nanopore. This liquid bridge then expands, becomes wider and wider, as the total entropy of the system increases, and finally takes over the system to yield the stable phase of high density. Applying this approach to the phenomenon of capillary evaporation similarly uncovers a multi-stage process. In this case, the layered structure of the metastable phase of high density is first destabilized by the adsorption of a few extra atoms. This is then followed by the nucleation of a bubble within the nanopore. Overall, the mechanism identified for capillary condensation is consistent with the scenario of Everett and Haynes~\cite{everett1972model} for a macroscopic system and with gauge cell Monte Carlo simulations by Neimark {\it et al.}~\cite{vishnyakov2003nucleation}. The results obtained here further establish the key role played by the nucleation of liquid bridges in the capillary condensation process, and of the nucleation of bubbles during capillary evaporation, and shed light on the interplay between the entropy and the structure of the adsorbed fluid throughout these processes.\\

{\bf Acknowledgements}
Partial funding for this research was provided by NSF through CAREER award DMR-1052808. Acknowledgement is made to the Donors of the American Chemical Society Petroleum Research Fund for partial support of this research through grant 548002-ND10.\\

\bibliography{JCP_S3}

\begin{thebibliography}{73}
\expandafter\ifx\csname natexlab\endcsname\relax\def\natexlab#1{#1}\fi
\expandafter\ifx\csname bibnamefont\endcsname\relax
  \def\bibnamefont#1{#1}\fi
\expandafter\ifx\csname bibfnamefont\endcsname\relax
  \def\bibfnamefont#1{#1}\fi
\expandafter\ifx\csname citenamefont\endcsname\relax
  \def\citenamefont#1{#1}\fi
\expandafter\ifx\csname url\endcsname\relax
  \def\url#1{\texttt{#1}}\fi
\expandafter\ifx\csname urlprefix\endcsname\relax\def\urlprefix{URL }\fi
\providecommand{\bibinfo}[2]{#2}
\providecommand{\eprint}[2][]{\url{#2}}

\bibitem[{\citenamefont{Derjaguin}(1992)}]{derjaguin1992theory}
\bibinfo{author}{\bibfnamefont{B.}~\bibnamefont{Derjaguin}},
  \bibinfo{journal}{Prog. Surf. Sci.} \textbf{\bibinfo{volume}{40}},
  \bibinfo{pages}{46} (\bibinfo{year}{1992}).

\bibitem[{\citenamefont{Coasne and Pellenq}(2004)}]{coasne2004grand}
\bibinfo{author}{\bibfnamefont{B.}~\bibnamefont{Coasne}} \bibnamefont{and}
  \bibinfo{author}{\bibfnamefont{R.-M.} \bibnamefont{Pellenq}},
  \bibinfo{journal}{J. Chem. Phys.}  (\bibinfo{year}{2004}).

\bibitem[{\citenamefont{Saugey et~al.}(2005)\citenamefont{Saugey, Bocquet, and
  Barrat}}]{saugey2005nucleation}
\bibinfo{author}{\bibfnamefont{A.}~\bibnamefont{Saugey}},
  \bibinfo{author}{\bibfnamefont{L.}~\bibnamefont{Bocquet}}, \bibnamefont{and}
  \bibinfo{author}{\bibfnamefont{J.}~\bibnamefont{Barrat}},
  \bibinfo{journal}{J. Phys. Chem. B} \textbf{\bibinfo{volume}{109}},
  \bibinfo{pages}{6520} (\bibinfo{year}{2005}).

\bibitem[{\citenamefont{Major et~al.}(2006)\citenamefont{Major, Houston,
  McGrath, Siepmann, and Zhu}}]{major2006viscous}
\bibinfo{author}{\bibfnamefont{R.}~\bibnamefont{Major}},
  \bibinfo{author}{\bibfnamefont{J.}~\bibnamefont{Houston}},
  \bibinfo{author}{\bibfnamefont{M.}~\bibnamefont{McGrath}},
  \bibinfo{author}{\bibfnamefont{J.}~\bibnamefont{Siepmann}}, \bibnamefont{and}
  \bibinfo{author}{\bibfnamefont{X.-Y.} \bibnamefont{Zhu}},
  \bibinfo{journal}{Phys. Rev. Lett.} \textbf{\bibinfo{volume}{96}},
  \bibinfo{pages}{177803} (\bibinfo{year}{2006}).

\bibitem[{\citenamefont{Casanova et~al.}(2007)\citenamefont{Casanova, Chiang,
  Li, and Schuller}}]{casanova2007direct}
\bibinfo{author}{\bibfnamefont{F.}~\bibnamefont{Casanova}},
  \bibinfo{author}{\bibfnamefont{C.~E.} \bibnamefont{Chiang}},
  \bibinfo{author}{\bibfnamefont{C.-P.} \bibnamefont{Li}}, \bibnamefont{and}
  \bibinfo{author}{\bibfnamefont{I.~K.} \bibnamefont{Schuller}},
  \bibinfo{journal}{Appl. Phys. Lett.} \textbf{\bibinfo{volume}{91}},
  \bibinfo{pages}{243103} (\bibinfo{year}{2007}).

\bibitem[{\citenamefont{Puibasset}(2008)}]{puibasset2008monte}
\bibinfo{author}{\bibfnamefont{J.}~\bibnamefont{Puibasset}},
  \bibinfo{journal}{Langmuir} \textbf{\bibinfo{volume}{25}},
  \bibinfo{pages}{903} (\bibinfo{year}{2008}).

\bibitem[{\citenamefont{Edison and Monson}(2013)}]{edison2013dynamics}
\bibinfo{author}{\bibfnamefont{J.~R.} \bibnamefont{Edison}} \bibnamefont{and}
  \bibinfo{author}{\bibfnamefont{P.~A.} \bibnamefont{Monson}},
  \bibinfo{journal}{J. Chem. Phys.} \textbf{\bibinfo{volume}{138}},
  \bibinfo{pages}{234709} (\bibinfo{year}{2013}).

\bibitem[{\citenamefont{Szoszkiewicz and
  Riedo}(2005)}]{szoszkiewicz2005nucleation}
\bibinfo{author}{\bibfnamefont{R.}~\bibnamefont{Szoszkiewicz}}
  \bibnamefont{and} \bibinfo{author}{\bibfnamefont{E.}~\bibnamefont{Riedo}},
  \bibinfo{journal}{Phys. Rev. Lett.} \textbf{\bibinfo{volume}{95}},
  \bibinfo{pages}{135502} (\bibinfo{year}{2005}).

\bibitem[{\citenamefont{Bruschi et~al.}(2010)\citenamefont{Bruschi, Mistura,
  Liu, Lee, Gsele, and Coasne}}]{bruschi2010capillary}
\bibinfo{author}{\bibfnamefont{L.}~\bibnamefont{Bruschi}},
  \bibinfo{author}{\bibfnamefont{G.}~\bibnamefont{Mistura}},
  \bibinfo{author}{\bibfnamefont{L.}~\bibnamefont{Liu}},
  \bibinfo{author}{\bibfnamefont{W.}~\bibnamefont{Lee}},
  \bibinfo{author}{\bibfnamefont{U.}~\bibnamefont{Gsele}}, \bibnamefont{and}
  \bibinfo{author}{\bibfnamefont{B.}~\bibnamefont{Coasne}},
  \bibinfo{journal}{Langmuir} \textbf{\bibinfo{volume}{26}},
  \bibinfo{pages}{11894} (\bibinfo{year}{2010}).

\bibitem[{\citenamefont{Greiner et~al.}(2010)\citenamefont{Greiner, Felts, Dai,
  King, and Carpick}}]{greiner2010local}
\bibinfo{author}{\bibfnamefont{C.}~\bibnamefont{Greiner}},
  \bibinfo{author}{\bibfnamefont{J.~R.} \bibnamefont{Felts}},
  \bibinfo{author}{\bibfnamefont{Z.}~\bibnamefont{Dai}},
  \bibinfo{author}{\bibfnamefont{W.~P.} \bibnamefont{King}}, \bibnamefont{and}
  \bibinfo{author}{\bibfnamefont{R.~W.} \bibnamefont{Carpick}},
  \bibinfo{journal}{Nano Lett.} \textbf{\bibinfo{volume}{10}},
  \bibinfo{pages}{4640} (\bibinfo{year}{2010}).

\bibitem[{\citenamefont{Lin et~al.}(1994)\citenamefont{Lin, Sinha, Drake, Wu,
  Thiyagarajan, and Stanley}}]{lin1994study}
\bibinfo{author}{\bibfnamefont{M.}~\bibnamefont{Lin}},
  \bibinfo{author}{\bibfnamefont{S.}~\bibnamefont{Sinha}},
  \bibinfo{author}{\bibfnamefont{J.}~\bibnamefont{Drake}},
  \bibinfo{author}{\bibfnamefont{X.-l.} \bibnamefont{Wu}},
  \bibinfo{author}{\bibfnamefont{P.}~\bibnamefont{Thiyagarajan}},
  \bibnamefont{and} \bibinfo{author}{\bibfnamefont{H.}~\bibnamefont{Stanley}},
  \bibinfo{journal}{Phys. Rev. Lett.} \textbf{\bibinfo{volume}{72}},
  \bibinfo{pages}{2207} (\bibinfo{year}{1994}).

\bibitem[{\citenamefont{Willett et~al.}(2000)\citenamefont{Willett, Adams,
  Johnson, and Seville}}]{willett2000capillary}
\bibinfo{author}{\bibfnamefont{C.~D.} \bibnamefont{Willett}},
  \bibinfo{author}{\bibfnamefont{M.~J.} \bibnamefont{Adams}},
  \bibinfo{author}{\bibfnamefont{S.~A.} \bibnamefont{Johnson}},
  \bibnamefont{and} \bibinfo{author}{\bibfnamefont{J.~P.}
  \bibnamefont{Seville}}, \bibinfo{journal}{Langmuir}
  \textbf{\bibinfo{volume}{16}}, \bibinfo{pages}{9396} (\bibinfo{year}{2000}).

\bibitem[{\citenamefont{Gogotsi et~al.}(2001)\citenamefont{Gogotsi, Libera,
  G{\"u}ven{\c{c}}-Yazicioglu, and Megaridis}}]{gogotsi2001situ}
\bibinfo{author}{\bibfnamefont{Y.}~\bibnamefont{Gogotsi}},
  \bibinfo{author}{\bibfnamefont{J.~A.} \bibnamefont{Libera}},
  \bibinfo{author}{\bibfnamefont{A.}~\bibnamefont{G{\"u}ven{\c{c}}-Yazicioglu}},
  \bibnamefont{and} \bibinfo{author}{\bibfnamefont{C.~M.}
  \bibnamefont{Megaridis}}, \bibinfo{journal}{Appl. Phys. Lett.}
  \textbf{\bibinfo{volume}{79}}, \bibinfo{pages}{1021} (\bibinfo{year}{2001}).

\bibitem[{\citenamefont{He et~al.}(2001)\citenamefont{He, Szuchmacher~Blum,
  Aston, Buenviaje, Overney, and Luginb{\"u}hl}}]{he2001critical}
\bibinfo{author}{\bibfnamefont{M.}~\bibnamefont{He}},
  \bibinfo{author}{\bibfnamefont{A.}~\bibnamefont{Szuchmacher~Blum}},
  \bibinfo{author}{\bibfnamefont{D.~E.} \bibnamefont{Aston}},
  \bibinfo{author}{\bibfnamefont{C.}~\bibnamefont{Buenviaje}},
  \bibinfo{author}{\bibfnamefont{R.~M.} \bibnamefont{Overney}},
  \bibnamefont{and}
  \bibinfo{author}{\bibfnamefont{R.}~\bibnamefont{Luginb{\"u}hl}},
  \bibinfo{journal}{J. Chem. Phys.} \textbf{\bibinfo{volume}{114}},
  \bibinfo{pages}{1355} (\bibinfo{year}{2001}).

\bibitem[{\citenamefont{Heuberger et~al.}(2001)\citenamefont{Heuberger,
  Z{\"a}ch, and Spencer}}]{heuberger2001density}
\bibinfo{author}{\bibfnamefont{M.}~\bibnamefont{Heuberger}},
  \bibinfo{author}{\bibfnamefont{M.}~\bibnamefont{Z{\"a}ch}}, \bibnamefont{and}
  \bibinfo{author}{\bibfnamefont{N.}~\bibnamefont{Spencer}},
  \bibinfo{journal}{Science} \textbf{\bibinfo{volume}{292}},
  \bibinfo{pages}{905} (\bibinfo{year}{2001}).

\bibitem[{\citenamefont{Patel et~al.}(2002)\citenamefont{Patel, Dodge,
  Alexander, Slobozhanin, Taylor, and Rosenblatt}}]{patel2002stability}
\bibinfo{author}{\bibfnamefont{N.~M.} \bibnamefont{Patel}},
  \bibinfo{author}{\bibfnamefont{M.~R.} \bibnamefont{Dodge}},
  \bibinfo{author}{\bibfnamefont{J.~I.~D.} \bibnamefont{Alexander}},
  \bibinfo{author}{\bibfnamefont{L.~A.} \bibnamefont{Slobozhanin}},
  \bibinfo{author}{\bibfnamefont{P.}~\bibnamefont{Taylor}}, \bibnamefont{and}
  \bibinfo{author}{\bibfnamefont{C.}~\bibnamefont{Rosenblatt}},
  \bibinfo{journal}{Phys. Rev. E} \textbf{\bibinfo{volume}{65}},
  \bibinfo{pages}{026306} (\bibinfo{year}{2002}).

\bibitem[{\citenamefont{Maeda and Israelachvili}(2002)}]{maeda2002nanoscale}
\bibinfo{author}{\bibfnamefont{N.}~\bibnamefont{Maeda}} \bibnamefont{and}
  \bibinfo{author}{\bibfnamefont{J.~N.} \bibnamefont{Israelachvili}},
  \bibinfo{journal}{J. Phys. Chem. B} \textbf{\bibinfo{volume}{106}},
  \bibinfo{pages}{3534} (\bibinfo{year}{2002}).

\bibitem[{\citenamefont{Jang et~al.}(2003)\citenamefont{Jang, Schatz, and
  Ratner}}]{jang2003capillary}
\bibinfo{author}{\bibfnamefont{J.}~\bibnamefont{Jang}},
  \bibinfo{author}{\bibfnamefont{G.~C.} \bibnamefont{Schatz}},
  \bibnamefont{and} \bibinfo{author}{\bibfnamefont{M.~A.}
  \bibnamefont{Ratner}}, \bibinfo{journal}{Phys. Rev. Lett.}
  \textbf{\bibinfo{volume}{90}}, \bibinfo{pages}{156104}
  (\bibinfo{year}{2003}).

\bibitem[{\citenamefont{Weeks et~al.}(2005)\citenamefont{Weeks, Vaughn, and
  DeYoreo}}]{weeks2005direct}
\bibinfo{author}{\bibfnamefont{B.~L.} \bibnamefont{Weeks}},
  \bibinfo{author}{\bibfnamefont{M.~W.} \bibnamefont{Vaughn}},
  \bibnamefont{and} \bibinfo{author}{\bibfnamefont{J.~J.}
  \bibnamefont{DeYoreo}}, \bibinfo{journal}{Langmuir}
  \textbf{\bibinfo{volume}{21}}, \bibinfo{pages}{8096} (\bibinfo{year}{2005}).

\bibitem[{\citenamefont{Berim and Ruckenstein}(2008)}]{berim2008nanodrop}
\bibinfo{author}{\bibfnamefont{G.~O.} \bibnamefont{Berim}} \bibnamefont{and}
  \bibinfo{author}{\bibfnamefont{E.}~\bibnamefont{Ruckenstein}},
  \bibinfo{journal}{J. Chem. Phys.} \textbf{\bibinfo{volume}{129}},
  \bibinfo{pages}{014708} (\bibinfo{year}{2008}).

\bibitem[{\citenamefont{Thommes et~al.}(2006)\citenamefont{Thommes, Smarsly,
  Groenewolt, Ravikovitch, and Neimark}}]{thommes2006adsorption}
\bibinfo{author}{\bibfnamefont{M.}~\bibnamefont{Thommes}},
  \bibinfo{author}{\bibfnamefont{B.}~\bibnamefont{Smarsly}},
  \bibinfo{author}{\bibfnamefont{M.}~\bibnamefont{Groenewolt}},
  \bibinfo{author}{\bibfnamefont{P.~I.} \bibnamefont{Ravikovitch}},
  \bibnamefont{and} \bibinfo{author}{\bibfnamefont{A.~V.}
  \bibnamefont{Neimark}}, \bibinfo{journal}{Langmuir}
  \textbf{\bibinfo{volume}{22}}, \bibinfo{pages}{756} (\bibinfo{year}{2006}).

\bibitem[{\citenamefont{Fortini and Dijkstra}(2006)}]{fortini2006phase}
\bibinfo{author}{\bibfnamefont{A.}~\bibnamefont{Fortini}} \bibnamefont{and}
  \bibinfo{author}{\bibfnamefont{M.}~\bibnamefont{Dijkstra}},
  \bibinfo{journal}{J. Phys. Condens. Matt.} \textbf{\bibinfo{volume}{18}},
  \bibinfo{pages}{L371} (\bibinfo{year}{2006}).

\bibitem[{\citenamefont{Horikawa et~al.}(2011)\citenamefont{Horikawa, Do, and
  Nicholson}}]{horikawa2011capillary}
\bibinfo{author}{\bibfnamefont{T.}~\bibnamefont{Horikawa}},
  \bibinfo{author}{\bibfnamefont{D.}~\bibnamefont{Do}}, \bibnamefont{and}
  \bibinfo{author}{\bibfnamefont{D.}~\bibnamefont{Nicholson}},
  \bibinfo{journal}{Adv. Colloid Interface Sci.}
  \textbf{\bibinfo{volume}{169}}, \bibinfo{pages}{40} (\bibinfo{year}{2011}).

\bibitem[{\citenamefont{Monson}(2012)}]{monson2012understanding}
\bibinfo{author}{\bibfnamefont{P.}~\bibnamefont{Monson}},
  \bibinfo{journal}{Microporous Mater.} \textbf{\bibinfo{volume}{160}},
  \bibinfo{pages}{47} (\bibinfo{year}{2012}).

\bibitem[{\citenamefont{Gommes}(2012)}]{gommes2012adsorption}
\bibinfo{author}{\bibfnamefont{C.~J.} \bibnamefont{Gommes}},
  \bibinfo{journal}{Langmuir} \textbf{\bibinfo{volume}{28}},
  \bibinfo{pages}{5101} (\bibinfo{year}{2012}).

\bibitem[{\citenamefont{Mszr et~al.}(2013)\citenamefont{Mszr, Hantal,
  Picaud, and Jedlovszky}}]{mszr2013adsorption}
\bibinfo{author}{\bibfnamefont{Z.~E.} \bibnamefont{Mszr}},
  \bibinfo{author}{\bibfnamefont{G.}~\bibnamefont{Hantal}},
  \bibinfo{author}{\bibfnamefont{S.}~\bibnamefont{Picaud}}, \bibnamefont{and}
  \bibinfo{author}{\bibfnamefont{P.}~\bibnamefont{Jedlovszky}},
  \bibinfo{journal}{J. Phys. Chem. C} \textbf{\bibinfo{volume}{117}},
  \bibinfo{pages}{6719} (\bibinfo{year}{2013}).

\bibitem[{\citenamefont{Zeng et~al.}(2014)\citenamefont{Zeng, Fan, Do, and
  Nicholson}}]{zeng2014condensation}
\bibinfo{author}{\bibfnamefont{Y.}~\bibnamefont{Zeng}},
  \bibinfo{author}{\bibfnamefont{C.}~\bibnamefont{Fan}},
  \bibinfo{author}{\bibfnamefont{D.}~\bibnamefont{Do}}, \bibnamefont{and}
  \bibinfo{author}{\bibfnamefont{D.}~\bibnamefont{Nicholson}},
  \bibinfo{journal}{J. Phys. Chem. C} \textbf{\bibinfo{volume}{118}},
  \bibinfo{pages}{3172} (\bibinfo{year}{2014}).

\bibitem[{\citenamefont{Hiratsuka et~al.}(2016)\citenamefont{Hiratsuka, Tanaka,
  and Miyahara}}]{hiratsuka2016mechanism}
\bibinfo{author}{\bibfnamefont{T.}~\bibnamefont{Hiratsuka}},
  \bibinfo{author}{\bibfnamefont{H.}~\bibnamefont{Tanaka}}, \bibnamefont{and}
  \bibinfo{author}{\bibfnamefont{M.~T.} \bibnamefont{Miyahara}},
  \bibinfo{journal}{ACS Nano}  (\bibinfo{year}{2016}).

\bibitem[{\citenamefont{Restagno et~al.}(2000)\citenamefont{Restagno, Bocquet,
  and Biben}}]{restagno2000metastability}
\bibinfo{author}{\bibfnamefont{F.}~\bibnamefont{Restagno}},
  \bibinfo{author}{\bibfnamefont{L.}~\bibnamefont{Bocquet}}, \bibnamefont{and}
  \bibinfo{author}{\bibfnamefont{T.}~\bibnamefont{Biben}},
  \bibinfo{journal}{Phys. Rev. Lett.} \textbf{\bibinfo{volume}{84}},
  \bibinfo{pages}{2433} (\bibinfo{year}{2000}).

\bibitem[{\citenamefont{Talanquer and Oxtoby}(2001)}]{talanquer2001nucleation}
\bibinfo{author}{\bibfnamefont{V.}~\bibnamefont{Talanquer}} \bibnamefont{and}
  \bibinfo{author}{\bibfnamefont{D.}~\bibnamefont{Oxtoby}},
  \bibinfo{journal}{J. Chem. Phys.} \textbf{\bibinfo{volume}{114}},
  \bibinfo{pages}{2793} (\bibinfo{year}{2001}).

\bibitem[{\citenamefont{Ustinov and Do}(2005)}]{ustinov2005modeling}
\bibinfo{author}{\bibfnamefont{E.}~\bibnamefont{Ustinov}} \bibnamefont{and}
  \bibinfo{author}{\bibfnamefont{D.}~\bibnamefont{Do}}, \bibinfo{journal}{J.
  Phys. Chem. B} \textbf{\bibinfo{volume}{109}}, \bibinfo{pages}{11653}
  (\bibinfo{year}{2005}).

\bibitem[{\citenamefont{Men et~al.}(2009)\citenamefont{Men, Yan, Jiang, Zhang,
  and Wang}}]{men2009nucleation}
\bibinfo{author}{\bibfnamefont{Y.}~\bibnamefont{Men}},
  \bibinfo{author}{\bibfnamefont{Q.}~\bibnamefont{Yan}},
  \bibinfo{author}{\bibfnamefont{G.}~\bibnamefont{Jiang}},
  \bibinfo{author}{\bibfnamefont{X.}~\bibnamefont{Zhang}}, \bibnamefont{and}
  \bibinfo{author}{\bibfnamefont{W.}~\bibnamefont{Wang}},
  \bibinfo{journal}{Phys. Rev. E} \textbf{\bibinfo{volume}{79}},
  \bibinfo{pages}{051602} (\bibinfo{year}{2009}).

\bibitem[{\citenamefont{Zhang and Chakrabarti}(1994)}]{zhang1994phase}
\bibinfo{author}{\bibfnamefont{Z.}~\bibnamefont{Zhang}} \bibnamefont{and}
  \bibinfo{author}{\bibfnamefont{A.}~\bibnamefont{Chakrabarti}},
  \bibinfo{journal}{Phys. Rev. E} \textbf{\bibinfo{volume}{50}},
  \bibinfo{pages}{R4290} (\bibinfo{year}{1994}).

\bibitem[{\citenamefont{Gelb and Gubbins}(1997)}]{gelb1997liquid}
\bibinfo{author}{\bibfnamefont{L.~D.} \bibnamefont{Gelb}} \bibnamefont{and}
  \bibinfo{author}{\bibfnamefont{K.}~\bibnamefont{Gubbins}},
  \bibinfo{journal}{Phys. Rev. E} \textbf{\bibinfo{volume}{56}},
  \bibinfo{pages}{3185} (\bibinfo{year}{1997}).

\bibitem[{\citenamefont{Landman}(1998)}]{landman1998nanotribological}
\bibinfo{author}{\bibfnamefont{U.}~\bibnamefont{Landman}},
  \bibinfo{journal}{Solid State Commun.} \textbf{\bibinfo{volume}{107}},
  \bibinfo{pages}{693} (\bibinfo{year}{1998}).

\bibitem[{\citenamefont{Leung and Luzar}(2000)}]{leung2000dynamics}
\bibinfo{author}{\bibfnamefont{K.}~\bibnamefont{Leung}} \bibnamefont{and}
  \bibinfo{author}{\bibfnamefont{A.}~\bibnamefont{Luzar}}, \bibinfo{journal}{J.
  Chem. Phys.} \textbf{\bibinfo{volume}{113}}, \bibinfo{pages}{5845}
  (\bibinfo{year}{2000}).

\bibitem[{\citenamefont{Yasuoka et~al.}(2000)\citenamefont{Yasuoka, Gao, and
  Zeng}}]{yasuoka2000molecular}
\bibinfo{author}{\bibfnamefont{K.}~\bibnamefont{Yasuoka}},
  \bibinfo{author}{\bibfnamefont{G.}~\bibnamefont{Gao}}, \bibnamefont{and}
  \bibinfo{author}{\bibfnamefont{X.~C.} \bibnamefont{Zeng}},
  \bibinfo{journal}{J. Chem. Phys.} \textbf{\bibinfo{volume}{112}},
  \bibinfo{pages}{4279} (\bibinfo{year}{2000}).

\bibitem[{\citenamefont{Liu and Grest}(1991)}]{liu1991wetting}
\bibinfo{author}{\bibfnamefont{A.~J.} \bibnamefont{Liu}} \bibnamefont{and}
  \bibinfo{author}{\bibfnamefont{G.~S.} \bibnamefont{Grest}},
  \bibinfo{journal}{Phys. Rev. A} \textbf{\bibinfo{volume}{44}},
  \bibinfo{pages}{R7894} (\bibinfo{year}{1991}).

\bibitem[{\citenamefont{Gac et~al.}(1994)\citenamefont{Gac, Patrykiejew, and
  Soko{\l}owski}}]{gac1994influence}
\bibinfo{author}{\bibfnamefont{W.}~\bibnamefont{Gac}},
  \bibinfo{author}{\bibfnamefont{A.}~\bibnamefont{Patrykiejew}},
  \bibnamefont{and}
  \bibinfo{author}{\bibfnamefont{S.}~\bibnamefont{Soko{\l}owski}},
  \bibinfo{journal}{Surf. Sci.} \textbf{\bibinfo{volume}{306}},
  \bibinfo{pages}{434} (\bibinfo{year}{1994}).

\bibitem[{\citenamefont{Gelb et~al.}(1999)\citenamefont{Gelb, Gubbins,
  Radhakrishnan, and Sliwinska-Bartkowiak}}]{gelb1999phase}
\bibinfo{author}{\bibfnamefont{L.~D.} \bibnamefont{Gelb}},
  \bibinfo{author}{\bibfnamefont{K.}~\bibnamefont{Gubbins}},
  \bibinfo{author}{\bibfnamefont{R.}~\bibnamefont{Radhakrishnan}},
  \bibnamefont{and}
  \bibinfo{author}{\bibfnamefont{M.}~\bibnamefont{Sliwinska-Bartkowiak}},
  \bibinfo{journal}{Rep. Prog. Phys.} \textbf{\bibinfo{volume}{62}},
  \bibinfo{pages}{1573} (\bibinfo{year}{1999}).

\bibitem[{\citenamefont{Bock and Schoen}(1999)}]{bock1999phase}
\bibinfo{author}{\bibfnamefont{H.}~\bibnamefont{Bock}} \bibnamefont{and}
  \bibinfo{author}{\bibfnamefont{M.}~\bibnamefont{Schoen}},
  \bibinfo{journal}{Phys. Rev. E} \textbf{\bibinfo{volume}{59}},
  \bibinfo{pages}{4122} (\bibinfo{year}{1999}).

\bibitem[{\citenamefont{Bolhuis and Chandler}(2000)}]{bolhuis2000transition}
\bibinfo{author}{\bibfnamefont{P.~G.} \bibnamefont{Bolhuis}} \bibnamefont{and}
  \bibinfo{author}{\bibfnamefont{D.}~\bibnamefont{Chandler}},
  \bibinfo{journal}{J. Chem. Phys.} \textbf{\bibinfo{volume}{113}},
  \bibinfo{pages}{8154} (\bibinfo{year}{2000}).

\bibitem[{\citenamefont{Stroud et~al.}(2001)\citenamefont{Stroud, Curry, and
  Cushman}}]{stroud2001capillary}
\bibinfo{author}{\bibfnamefont{W.~J.} \bibnamefont{Stroud}},
  \bibinfo{author}{\bibfnamefont{J.~E.} \bibnamefont{Curry}}, \bibnamefont{and}
  \bibinfo{author}{\bibfnamefont{J.~H.} \bibnamefont{Cushman}},
  \bibinfo{journal}{Langmuir} \textbf{\bibinfo{volume}{17}},
  \bibinfo{pages}{688} (\bibinfo{year}{2001}).

\bibitem[{\citenamefont{Liu and Monson}(2006)}]{liu2006monte}
\bibinfo{author}{\bibfnamefont{J.-C.} \bibnamefont{Liu}} \bibnamefont{and}
  \bibinfo{author}{\bibfnamefont{P.}~\bibnamefont{Monson}},
  \bibinfo{journal}{Ind. Eng. Chem. Res.} \textbf{\bibinfo{volume}{45}},
  \bibinfo{pages}{5649} (\bibinfo{year}{2006}).

\bibitem[{\citenamefont{Mota and Esteves}(2007)}]{mota2007simplified}
\bibinfo{author}{\bibfnamefont{J.~P.} \bibnamefont{Mota}} \bibnamefont{and}
  \bibinfo{author}{\bibfnamefont{I.~A.} \bibnamefont{Esteves}},
  \bibinfo{journal}{Adsorption} \textbf{\bibinfo{volume}{13}},
  \bibinfo{pages}{21} (\bibinfo{year}{2007}).

\bibitem[{\citenamefont{Winkler et~al.}(2010)\citenamefont{Winkler, Wilms,
  Virnau, and Binder}}]{winkler2010capillary}
\bibinfo{author}{\bibfnamefont{A.}~\bibnamefont{Winkler}},
  \bibinfo{author}{\bibfnamefont{D.}~\bibnamefont{Wilms}},
  \bibinfo{author}{\bibfnamefont{P.}~\bibnamefont{Virnau}}, \bibnamefont{and}
  \bibinfo{author}{\bibfnamefont{K.}~\bibnamefont{Binder}},
  \bibinfo{journal}{J. Chem. Phys.} \textbf{\bibinfo{volume}{133}},
  \bibinfo{pages}{164702} (\bibinfo{year}{2010}).

\bibitem[{\citenamefont{Nguyen et~al.}(2011)\citenamefont{Nguyen, Do, and
  Nicholson}}]{nguyen2011monte}
\bibinfo{author}{\bibfnamefont{V.~T.} \bibnamefont{Nguyen}},
  \bibinfo{author}{\bibfnamefont{D.}~\bibnamefont{Do}}, \bibnamefont{and}
  \bibinfo{author}{\bibfnamefont{D.}~\bibnamefont{Nicholson}},
  \bibinfo{journal}{J. Phys. Chem. B} \textbf{\bibinfo{volume}{115}},
  \bibinfo{pages}{7862} (\bibinfo{year}{2011}).

\bibitem[{\citenamefont{Gor et~al.}(2012)\citenamefont{Gor, Rasmussen, and
  Neimark}}]{gor2012capillary}
\bibinfo{author}{\bibfnamefont{G.~Y.} \bibnamefont{Gor}},
  \bibinfo{author}{\bibfnamefont{C.~J.} \bibnamefont{Rasmussen}},
  \bibnamefont{and} \bibinfo{author}{\bibfnamefont{A.~V.}
  \bibnamefont{Neimark}}, \bibinfo{journal}{Langmuir}
  \textbf{\bibinfo{volume}{28}}, \bibinfo{pages}{12100} (\bibinfo{year}{2012}).

\bibitem[{\citenamefont{Siderius and Shen}(2013)}]{siderius2013use}
\bibinfo{author}{\bibfnamefont{D.~W.} \bibnamefont{Siderius}} \bibnamefont{and}
  \bibinfo{author}{\bibfnamefont{V.~K.} \bibnamefont{Shen}},
  \bibinfo{journal}{J. Phys. Chem. C} \textbf{\bibinfo{volume}{117}},
  \bibinfo{pages}{5861} (\bibinfo{year}{2013}).

\bibitem[{\citenamefont{van Meel et~al.}(2015)\citenamefont{van Meel, Liu, and
  Frenkel}}]{van2015mechanism}
\bibinfo{author}{\bibfnamefont{J.}~\bibnamefont{van Meel}},
  \bibinfo{author}{\bibfnamefont{Y.}~\bibnamefont{Liu}}, \bibnamefont{and}
  \bibinfo{author}{\bibfnamefont{D.}~\bibnamefont{Frenkel}},
  \bibinfo{journal}{Mol. Phys.} \textbf{\bibinfo{volume}{113}},
  \bibinfo{pages}{2742} (\bibinfo{year}{2015}).

\bibitem[{\citenamefont{Kornev et~al.}(2002)\citenamefont{Kornev, Shingareva,
  and Neimark}}]{kornev2002capillary}
\bibinfo{author}{\bibfnamefont{K.~G.} \bibnamefont{Kornev}},
  \bibinfo{author}{\bibfnamefont{I.~K.} \bibnamefont{Shingareva}},
  \bibnamefont{and} \bibinfo{author}{\bibfnamefont{A.~V.}
  \bibnamefont{Neimark}}, \bibinfo{journal}{Adv. Colloid Interface Sci.}
  \textbf{\bibinfo{volume}{96}}, \bibinfo{pages}{143} (\bibinfo{year}{2002}).

\bibitem[{\citenamefont{Neimark et~al.}(2003)\citenamefont{Neimark,
  Ravikovitch, and Vishnyakov}}]{neimark2003bridging}
\bibinfo{author}{\bibfnamefont{A.~V.} \bibnamefont{Neimark}},
  \bibinfo{author}{\bibfnamefont{P.~I.} \bibnamefont{Ravikovitch}},
  \bibnamefont{and}
  \bibinfo{author}{\bibfnamefont{A.}~\bibnamefont{Vishnyakov}},
  \bibinfo{journal}{J. Phys. Condens. Matt.} \textbf{\bibinfo{volume}{15}},
  \bibinfo{pages}{347} (\bibinfo{year}{2003}).

\bibitem[{\citenamefont{Vishnyakov and
  Neimark}(2003{\natexlab{a}})}]{vishnyakov2003monte}
\bibinfo{author}{\bibfnamefont{A.}~\bibnamefont{Vishnyakov}} \bibnamefont{and}
  \bibinfo{author}{\bibfnamefont{A.~V.} \bibnamefont{Neimark}},
  \bibinfo{journal}{Langmuir} \textbf{\bibinfo{volume}{19}},
  \bibinfo{pages}{3240} (\bibinfo{year}{2003}{\natexlab{a}}).

\bibitem[{\citenamefont{Vishnyakov and
  Neimark}(2003{\natexlab{b}})}]{vishnyakov2003nucleation}
\bibinfo{author}{\bibfnamefont{A.}~\bibnamefont{Vishnyakov}} \bibnamefont{and}
  \bibinfo{author}{\bibfnamefont{A.~V.} \bibnamefont{Neimark}},
  \bibinfo{journal}{J. Chem. Phys.} \textbf{\bibinfo{volume}{119}},
  \bibinfo{pages}{9755} (\bibinfo{year}{2003}{\natexlab{b}}).

\bibitem[{\citenamefont{Neimark and
  Vishnyakov}(2005{\natexlab{a}})}]{neimark2005monte}
\bibinfo{author}{\bibfnamefont{A.~V.} \bibnamefont{Neimark}} \bibnamefont{and}
  \bibinfo{author}{\bibfnamefont{A.}~\bibnamefont{Vishnyakov}},
  \bibinfo{journal}{J. Chem. Phys.} \textbf{\bibinfo{volume}{122}},
  \bibinfo{pages}{174508} (\bibinfo{year}{2005}{\natexlab{a}}).

\bibitem[{\citenamefont{Neimark and
  Vishnyakov}(2005{\natexlab{b}})}]{neimark2005birth}
\bibinfo{author}{\bibfnamefont{A.~V.} \bibnamefont{Neimark}} \bibnamefont{and}
  \bibinfo{author}{\bibfnamefont{A.}~\bibnamefont{Vishnyakov}},
  \bibinfo{journal}{J. Chem. Phys.} \textbf{\bibinfo{volume}{122}},
  \bibinfo{pages}{054707} (\bibinfo{year}{2005}{\natexlab{b}}).

\bibitem[{\citenamefont{Neimark and
  Vishnyakov}(2005{\natexlab{c}})}]{neimark2005vapor}
\bibinfo{author}{\bibfnamefont{A.~V.} \bibnamefont{Neimark}} \bibnamefont{and}
  \bibinfo{author}{\bibfnamefont{A.}~\bibnamefont{Vishnyakov}},
  \bibinfo{journal}{J. Phys. Chem. B} \textbf{\bibinfo{volume}{109}},
  \bibinfo{pages}{5962} (\bibinfo{year}{2005}{\natexlab{c}}).

\bibitem[{\citenamefont{Everett and Haynes}(1972)}]{everett1972model}
\bibinfo{author}{\bibfnamefont{D.}~\bibnamefont{Everett}} \bibnamefont{and}
  \bibinfo{author}{\bibfnamefont{J.}~\bibnamefont{Haynes}},
  \bibinfo{journal}{J. Colloid Interface Sci.} \textbf{\bibinfo{volume}{38}},
  \bibinfo{pages}{125} (\bibinfo{year}{1972}).

\bibitem[{\citenamefont{Oxtoby and Evans}(1988)}]{oxtoby1988nonclassical}
\bibinfo{author}{\bibfnamefont{D.~W.} \bibnamefont{Oxtoby}} \bibnamefont{and}
  \bibinfo{author}{\bibfnamefont{R.}~\bibnamefont{Evans}}, \bibinfo{journal}{J.
  Chem. Phys.} \textbf{\bibinfo{volume}{89}}, \bibinfo{pages}{7521}
  (\bibinfo{year}{1988}).

\bibitem[{\citenamefont{ten Wolde and Frenkel}(1998)}]{ten1998numerical}
\bibinfo{author}{\bibfnamefont{P.~R.} \bibnamefont{ten Wolde}}
  \bibnamefont{and} \bibinfo{author}{\bibfnamefont{D.}~\bibnamefont{Frenkel}},
  \bibinfo{journal}{J. Chem. Phys.} \textbf{\bibinfo{volume}{109}},
  \bibinfo{pages}{9919} (\bibinfo{year}{1998}).

\bibitem[{\citenamefont{Ten~Wolde et~al.}(1999)\citenamefont{Ten~Wolde,
  Ruiz-Montero, and Frenkel}}]{ten1999numerical}
\bibinfo{author}{\bibfnamefont{P.~R.} \bibnamefont{Ten~Wolde}},
  \bibinfo{author}{\bibfnamefont{M.~J.} \bibnamefont{Ruiz-Montero}},
  \bibnamefont{and} \bibinfo{author}{\bibfnamefont{D.}~\bibnamefont{Frenkel}},
  \bibinfo{journal}{J. Chem. Phys.} \textbf{\bibinfo{volume}{110}},
  \bibinfo{pages}{1591} (\bibinfo{year}{1999}).

\bibitem[{\citenamefont{Shen and Debenedetti}(1999)}]{shen1999computational}
\bibinfo{author}{\bibfnamefont{V.~K.} \bibnamefont{Shen}} \bibnamefont{and}
  \bibinfo{author}{\bibfnamefont{P.~G.} \bibnamefont{Debenedetti}},
  \bibinfo{journal}{J. Chem. Phys.} \textbf{\bibinfo{volume}{111}},
  \bibinfo{pages}{3581} (\bibinfo{year}{1999}).

\bibitem[{\citenamefont{Desgranges and Delhommelle}(2011)}]{desgranges2011role}
\bibinfo{author}{\bibfnamefont{C.}~\bibnamefont{Desgranges}} \bibnamefont{and}
  \bibinfo{author}{\bibfnamefont{J.}~\bibnamefont{Delhommelle}},
  \bibinfo{journal}{J. Am. Chem. Soc.} \textbf{\bibinfo{volume}{133}},
  \bibinfo{pages}{2872} (\bibinfo{year}{2011}).

\bibitem[{\citenamefont{Patel et~al.}(2011)\citenamefont{Patel, Varilly,
  Chandler, and Garde}}]{patel2011quantifying}
\bibinfo{author}{\bibfnamefont{A.~J.} \bibnamefont{Patel}},
  \bibinfo{author}{\bibfnamefont{P.}~\bibnamefont{Varilly}},
  \bibinfo{author}{\bibfnamefont{D.}~\bibnamefont{Chandler}}, \bibnamefont{and}
  \bibinfo{author}{\bibfnamefont{S.}~\bibnamefont{Garde}}, \bibinfo{journal}{J.
  Stat. Phys.} \textbf{\bibinfo{volume}{145}}, \bibinfo{pages}{265}
  (\bibinfo{year}{2011}).

\bibitem[{\citenamefont{Desgranges and
  Delhommelle}(2014)}]{desgranges2014unraveling}
\bibinfo{author}{\bibfnamefont{C.}~\bibnamefont{Desgranges}} \bibnamefont{and}
  \bibinfo{author}{\bibfnamefont{J.}~\bibnamefont{Delhommelle}},
  \bibinfo{journal}{J. Am. Chem. Soc.} \textbf{\bibinfo{volume}{136}},
  \bibinfo{pages}{8145} (\bibinfo{year}{2014}).

\bibitem[{\citenamefont{Lauricella et~al.}(2015)\citenamefont{Lauricella,
  Meloni, Liang, English, Kusalik, and Ciccotti}}]{lauricella2015clathrate}
\bibinfo{author}{\bibfnamefont{M.}~\bibnamefont{Lauricella}},
  \bibinfo{author}{\bibfnamefont{S.}~\bibnamefont{Meloni}},
  \bibinfo{author}{\bibfnamefont{S.}~\bibnamefont{Liang}},
  \bibinfo{author}{\bibfnamefont{N.~J.} \bibnamefont{English}},
  \bibinfo{author}{\bibfnamefont{P.~G.} \bibnamefont{Kusalik}},
  \bibnamefont{and} \bibinfo{author}{\bibfnamefont{G.}~\bibnamefont{Ciccotti}},
  \bibinfo{journal}{J. Chem. Phys.} \textbf{\bibinfo{volume}{142}},
  \bibinfo{pages}{244503} (\bibinfo{year}{2015}).

\bibitem[{\citenamefont{Desgranges and Delhommelle}(2016{\natexlab{a}})}]{FS1}
\bibinfo{author}{\bibfnamefont{C.}~\bibnamefont{Desgranges}} \bibnamefont{and}
  \bibinfo{author}{\bibfnamefont{J.}~\bibnamefont{Delhommelle}},
  \bibinfo{journal}{J. Chem. Phys.} \textbf{\bibinfo{volume}{145}},
  \bibinfo{pages}{204112} (\bibinfo{year}{2016}{\natexlab{a}}).

\bibitem[{\citenamefont{Desgranges and Delhommelle}(2016{\natexlab{b}})}]{FS2}
\bibinfo{author}{\bibfnamefont{C.}~\bibnamefont{Desgranges}} \bibnamefont{and}
  \bibinfo{author}{\bibfnamefont{J.}~\bibnamefont{Delhommelle}},
  \bibinfo{journal}{J. Chem. Phys.} \textbf{\bibinfo{volume}{145}},
  \bibinfo{pages}{234505} (\bibinfo{year}{2016}{\natexlab{b}}).

\bibitem[{\citenamefont{Torrie and Valleau}(1977)}]{torrie1977nonphysical}
\bibinfo{author}{\bibfnamefont{G.~M.} \bibnamefont{Torrie}} \bibnamefont{and}
  \bibinfo{author}{\bibfnamefont{J.~P.} \bibnamefont{Valleau}},
  \bibinfo{journal}{J. Comput. Phys.} \textbf{\bibinfo{volume}{23}},
  \bibinfo{pages}{187} (\bibinfo{year}{1977}).

\bibitem[{\citenamefont{Ravikovitch et~al.}(1997)\citenamefont{Ravikovitch,
  Wei, Chueh, Haller, and Neimark}}]{ravikovitch1997evaluation}
\bibinfo{author}{\bibfnamefont{P.}~\bibnamefont{Ravikovitch}},
  \bibinfo{author}{\bibfnamefont{D.}~\bibnamefont{Wei}},
  \bibinfo{author}{\bibfnamefont{W.}~\bibnamefont{Chueh}},
  \bibinfo{author}{\bibfnamefont{G.}~\bibnamefont{Haller}}, \bibnamefont{and}
  \bibinfo{author}{\bibfnamefont{A.}~\bibnamefont{Neimark}},
  \bibinfo{journal}{J. Phys. Chem. B} \textbf{\bibinfo{volume}{101}},
  \bibinfo{pages}{3671} (\bibinfo{year}{1997}).

\bibitem[{\citenamefont{Allen and Tildesley}(1987)}]{Allen}
\bibinfo{author}{\bibfnamefont{M.~P.} \bibnamefont{Allen}} \bibnamefont{and}
  \bibinfo{author}{\bibfnamefont{D.~J.} \bibnamefont{Tildesley}},
  \emph{\bibinfo{title}{Computer Simulation of Liquids}}
  (\bibinfo{publisher}{Clarendon, Oxford}, \bibinfo{year}{1987}).

\bibitem[{\citenamefont{Waghe et~al.}(2012)\citenamefont{Waghe, Rasaiah, and
  Hummer}}]{Waghe}
\bibinfo{author}{\bibfnamefont{A.}~\bibnamefont{Waghe}},
  \bibinfo{author}{\bibfnamefont{J.~C.~R.} \bibnamefont{Rasaiah}},
  \bibnamefont{and} \bibinfo{author}{\bibfnamefont{G.}~\bibnamefont{Hummer}},
  \bibinfo{journal}{J. Chem. Phys.} \textbf{\bibinfo{volume}{137}},
  \bibinfo{pages}{044709} (\bibinfo{year}{2012}).

\bibitem[{\citenamefont{Peterson and Gubbins}(1987)}]{peterson1987phase}
\bibinfo{author}{\bibfnamefont{B.~K.} \bibnamefont{Peterson}} \bibnamefont{and}
  \bibinfo{author}{\bibfnamefont{K.~E.} \bibnamefont{Gubbins}},
  \bibinfo{journal}{Mol. Phys.} \textbf{\bibinfo{volume}{62}},
  \bibinfo{pages}{215} (\bibinfo{year}{1987}).

\end{thebibliography}

\end{document}